\documentclass[11pt,a4paper]{article}

\usepackage{jheppub}

\usepackage{graphicx}
\usepackage{bm}
\usepackage{dcolumn}
\usepackage{color}
\usepackage{pst-plot}

\newcommand{\beq}{\begin{eqnarray}}
\newcommand{\eeq}{\end{eqnarray}}

\title {Hawking emission from quantum gravity black holes}
\author[a,b]{Piero Nicolini}
\author[c]{Elizabeth Winstanley}
\affiliation[a]{Frankfurt Institute for Advanced Studies (FIAS), 
Frankfurt am Main, Germany}
\affiliation[b]{
Institut f\"ur Theoretische Physik,  J. W.
Goethe-Universit\"at,
Frankfurt am Main, Germany}
\affiliation[c]{School of Mathematics and Statistics, The University of Sheffield,
Hicks Building, Hounsfield Road, Sheffield. S3 7RH United Kingdom}

\emailAdd{nicolini@th.physik.uni-frankfurt.de}
\emailAdd{E.Winstanley@sheffield.ac.uk}

\abstract{
We address the issue of modelling quantum gravity effects in the evaporation of higher dimensional black holes in order to go beyond the usual semi-classical approximation. After reviewing the existing six families of quantum gravity corrected black hole geometries, we focus our work on non-commutative geometry inspired black holes, which encode model independent characteristics, are unaffected by the quantum back reaction and have an analytical form compact enough for numerical simulations. We consider the higher dimensional, spherically symmetric case and we proceed with a complete analysis of the brane/bulk emission for scalar fields. The key feature which makes the evaporation of non-commutative black holes so peculiar is the possibility of having a maximum temperature. Contrary to what happens with classical Schwarzschild black holes, the emission is dominated by low frequency field modes on the brane. This is a distinctive and potentially testable signature which might disclose further features about the nature of quantum gravity.
}

\begin{document}

\maketitle
\flushbottom

\section{Introduction}
\label{sec:intro}

\begin{figure}
\begin{center}
\includegraphics[width=10.0cm]{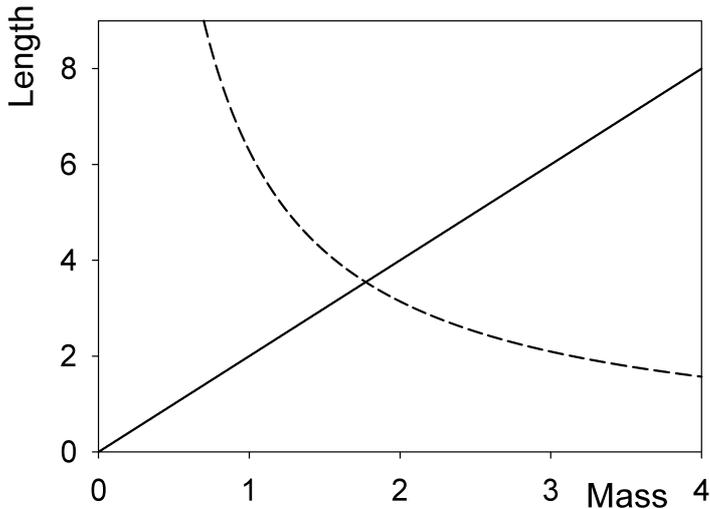}
\caption{\label{fig:bhformation} Particle Compton wavelengths (dotted curve) and horizon radii (solid line) as a function of the mass in Planck units. The intersection of the two curves corresponds to the formation of a mini black hole. }
\end{center}
\end{figure}

The possibility for a black hole to emit thermal radiation like a black body, often called black hole evaporation, is the first and maybe one of the best-known results of the combination of quantum field theory with general relativity.  Black hole evaporation is a topic of primary importance in fundamental physics, since it affects many research areas, spanning thermodynamics, relativity and particle physics. In addition, black hole evaporation represents the first convincing insight into a possible theory of quantum gravity.  However, despite the fact that the original derivation due to Hawking is dated back to 1975 \cite{Haw75} we do not yet have direct evidence about the actual observation of this phenomenon. Astrophysical black holes behave like classical objects due to their large mass. On the other hand, for microscopic black holes the evaporation is expected to be relevant. For black hole masses around $M\sim 10^{-11}$ kg, we have temperatures about $T\sim 10^{12}$ K and horizon radii about $r_h\sim 10^{-16}$~m. These are typical parameters of primordial black holes, black holes that might have formed due to the high density fluctuations of the early universe. Being extremely bright, their detection is expected at the Fermi Gamma-ray Space Telescope \cite{GLAST}. For even smaller sized black holes, we fully enter the regime of particle physics and we need an increased degree of compression of matter to create a mini black hole. According to the hoop conjecture, a ``particle black hole'' would form if its Compton wavelength equals the corresponding horizon radius \cite{Tho72,Meade:2007sz} (see figure~\ref{fig:bhformation}). This implies that mini black holes must have masses of the order of the Planck mass, $M\sim M_P$, and radii of the order of the Planck length, $r_h\sim L_P$, a fact that creates formidable problems \cite{DFG10,Spa11}: on the experimental side the Planck scale is about 15 orders of magnitude higher than the scale of current high energy physics experiments, while on the theoretical side we do not have yet a full formulation of quantum gravity which is suitable for efficiently describing evaporating black holes. This puzzling situation has no concrete ways out unless we make further hypotheses. If the space-time is endowed with additional spatial dimensions, it is possible to lower the fundamental scale of quantum gravity to an energy scale accessible to current particle physics experiments, namely $M_\star\sim 1\ \mathrm{TeV}$ \cite{AADD98,ADD99,RaS99}. A lower fundamental scale of quantum gravity implies a stronger gravitational interaction which allows the gravitational collapse of matter compressed  at distances of the order of $10^{-4}$ fermi \cite{DiL01}, i.e.~the typical length scales under scrutiny at the LHC \cite{CMS11}. This fascinating opportunity has led to intensive research activity whose main results can be found in various reviews, see, for example, \cite{Cav03,Kan04,Hos05,Web05,CaS06,Win07,BlN10,Kanti:2008eq,Kanti:2009sz}.

Despite these efforts and the large number of papers published in the field, the theoretical scenario is still uncertain. For instance, the lower quantum-gravitational energy scale requires higher dimensional metrics that in the case of  charged rotating black holes are not known analytically \cite{Allahverdizadeh:2010hi}. Even when we have analytic space-time geometries for describing some phases of the life of a microscopic black hole, the master equations for the propagation of matter fields  can be integrated only via accurate numerical methods (see, for example, \cite{BCC06,CKW06,CDK07,CDK08,Harris:2003eg}). Finally, we ignore the Planck phase, namely the fate of the black hole in the terminal phase of the evaporation, when its temperature equals the fundamental scale $T\sim M\sim M_\star$. We recall that evaporating black holes are conventionally described in terms of semi-classical gravity, which is valid only if the black hole metric is not modified by the emitted particles, i.e.~if $T\ll M$.

In the absence of a viable description of the Planck phase by some quantum theory of gravity, there have been several attempts to incorporate one or more features we expect from  quantum gravity in the formalism of the evaporation by means of effective theories. According to the formalism adopted, one has to deal with features like asymptotic freedom, non-commutative character, minimal area, minimal length, or non-locality, but in the end the crucial aspect in all cases is the possibility for the space-time to undergo a transition from a smooth differentiable manifold to a fractal surface, plagued by quantum uncertainty and the loss of resolution \cite{AAS97,AJL05,LaR05,Mod09,MoN10a,NiS11,NiN11,MuS11,Cal11a,Cal11b}. As a result, the corresponding quantum gravity modified black hole metrics have, in most cases, an equivalent qualitative behavior: the prevalent scenario is that of short-scale regularized metrics and the possibility of horizon extremization even for the neutral, static case, with a consequent cooling phase towards a zero temperature remnant as the mass approaches $M\sim M_\star$ \cite{Sca99,BoR99,Mod04,Nic09,MMN11}.  There is an additional advantage: since these quantum gravity black holes are significantly colder than the corresponding classical black holes, throughout the evaporation their metrics are not affected by a significant back-reaction. The metric modifications are already taken into account by the quantum-gravity corrections to the usual background geometries. As a consequence, one can safely use quantum field theory in curved space to study the evaporation of these black holes, without a breakdown of the formalism.

Given this background, it is imperative to study the evaporation of quantum gravity black holes by performing a detailed analysis of the brane/bulk emission, including the grey-body factors. This would be the first step in the quest for consistent signatures of evaporating black holes which are a requirement for starting any study of quantum gravity phenomenology.

The outline of this paper is as follows.
In section \ref{sec:QGBHs} we briefly review the existing six families of quantum gravity corrected black hole geometries,
before focussing our attention on neutral static, spherically symmetric non-commutative black holes.  We outline the key features of the metric and temperature of these black holes, in particular showing that the back-reaction of quantum fields on these
geometries is negligibly small, both for black holes potentially created at the LHC and in cosmic rays.
Hawking radiation of scalar particles, both on the brane and in the bulk, is studied in detail in section \ref{sec:Hawkrad}.
As well as the conventional fluxes of particles and energy, we also introduce an emission spectrum modified by non-commutative effects, although the latter does not give significantly different results.
We present our conclusions in section \ref{sec:conc}.

  \section{Quantum gravity black holes}
\label{sec:QGBHs}

For sake of clarity, we shall classify the quantum gravity corrected metrics currently available in the literature into six families, according to the mechanisms used to obtain the modifications with respect to classical space-times. They include non-local gravity black holes \cite{MMN11,Modesto11}, non-commutative geometry inspired black holes (NCBHs) \cite{Nic05,NSS06a,NSS06b,MKP07b,ANS07,BMS08,BMM09,ABN09,NiS10,SmS10,ABN10,BGM10,MuN11,NiT11},   generalized uncertainty principle black holes \cite{ACS01,AAL06,MKP07a,CMP11}, loop quantum black holes (LQBHs) \cite{Mod06,Mod08,MoP09,Mod10,HMP10,CaM10},  asymptotically safe gravity black holes (ASGBHs) \cite{BoR00,BoR06,BuK10,FLR10,CaE10} and a generic category of short scale modified metrics \cite{MbK05,Myu09,Par09,KiK10,GHS10,Nic10,MuS10} (for a review of earlier contributions see \cite{Ansoldi:2008jw}).

As a first paper in the area, we start our analysis from the most simple case, namely the neutral, spherically
symmetric static black hole, postponing the study of axisymmetric geometries to the future. As a second point, we will not study all the existing geometries mentioned above, but just the case of NCBHs. This choice is motivated by the following reasons. NCBHs are the richest family of quantum gravity improved black hole space-times.
There exist higher-dimensional static \cite{Riz06}, charged \cite{SSN09}, rotating \cite{SmS10} and charged rotating \cite{MoN10b} NCBHs, the latter only for low angular momenta as is the case for classical black holes. Therefore NCBHs are the only ones that can currently provide a complete scenario and it is worth starting from them in view of future investigations.
In addition, NCBHs have been found to be a sub-class of non-local gravity modified space-times \cite{MMN11,Modesto11}. As a consequence, NCBHs encode features that are common to more than one formulation and might lead to model-independent results.
As a side motivation, the analytic form of NCBHs is compact enough to implement into numerical simulations, as has already been done in previous contributions \cite{CaN08} (without studying the details of the Hawking emission), including in the development of a Monte Carlo event generator \cite{Gin10}.

We now proceed by recalling some basic facts about NCBHs. Non-commutative geometry is an old idea,  concerning the possibility that co-ordinate operators might fail to commute in some extreme energy limit \cite{Sny47a}.
This opened up a vast research area, with a high degree of mathematical sophistication (for an incomplete list of reviews on the topic see \cite{Lan97,Mad00,DoN01,Sza03}).
Despite the huge literature in the field, a formulation of the non-commutative equivalent of general relativity is still missing. The best one can do is to consider the average effect of non-commutative fluctuations and study the consequences for the gravity field equations. As a result, one can incorporate the presence of non-commutative effects by a non-standard energy-momentum tensor, while keeping  the Einstein tensor formally unchanged. It turns out that for the specific case of a static spherically symmetric source,  the usual point-like profile is no longer physically meaningful and must be replaced by a Gaussian distribution
\begin{equation}
T_0^{\ 0} (\vec{x})=-M\delta(\vec{x})\to \frac{M}{(4\pi\theta)^{\frac{n+3}{2}}}
e^{-\vec{x}^2/4\theta} ,
\end{equation}
where $n$ is the number of extra dimensions and $\theta$ the non-commutative parameter with dimensions of a length squared, that encodes a minimal length in the manifold. This is a key result which has been derived both within non-commutative geometry \cite{SmS04,SSN06,BGM10}  and non-local gravity \cite{MMN11}.
Covariant conservation and the additional condition $g_{00}=-1/g_{11}$ completely specify the energy momentum tensor, which then generates the solution
\begin{equation}
ds^{2}= -h(r) \, dt^{2}+h(r)^{-1} \, dr^{2} + r^{2}  \, d\Omega _{n+2}^{2}
\label{eq:metric}
\end{equation}
with
\begin{equation}
h(r) = 1 - \frac{1}{r^{n+1}}\left(\frac{1}{M_\star \sqrt{\pi}}\right)^{n+1} \left(\frac{M}{M_\star}\right)\left[\frac{ 8 \gamma \left( \frac {n+3}{2}, \frac {r^{2}}{4\theta} \right)}{n+2}\right] ,
\label{eq:hr}
\end{equation}
where $d\Omega _{n+2}^{2}$ is the metric of the $(n+2)$ dimensional unit sphere
\begin{equation}
d\Omega_{n+2}^{2}
=d\vartheta_{n+1}^2+\sin^2\vartheta_{n+1}\left(d\vartheta_n^2+\sin^2\vartheta_n^2\left(\dots +
\sin^2\vartheta_2(d\vartheta_1^2+\sin^2\vartheta_1d\varphi^2)\dots\right)\right)
\end{equation}
and
\begin{equation}
\gamma \left( \frac {n+3}{2}, \frac {r^{2}}{4\theta} \right)=\int_0^{r^2/4\theta}dt \ t^{\frac{n+1}{2}}\ e^{-t }
\label{eq:gamma}
\end{equation}
is the incomplete Euler gamma function. In the above,
the angles are defined as $0 < \varphi < 2\pi$ and $0 < \vartheta_i < \pi$, for $i = 1, \dots, n + 1$, while
the minimal length $\sqrt{\theta}$ is not set \textit{a priori}. However it is reasonable to have
$\sqrt{\theta}\sim M_\star^{-1}\sim 10^{-4}$ fermi, where the fundamental scale of quantum gravity is
\begin{equation}
M_\star \sim \left(\frac{L_P}{R}\right)^{\frac{n}{n+2}} M_P\sim 1 \  \mathrm{TeV},
\end{equation}
with $R$ the size of the extra dimensions.

The above line element (\ref{eq:metric}) approaches the usual higher dimensional
Schwarzschild solution for large radii, namely $r\gg \sqrt{\theta}$, where we expect quantum gravity corrections to be negligible.
Conversely, for small radii $r\lesssim \sqrt{\theta}$, the line element (\ref{eq:metric}) approaches a local de Sitter core
\begin{equation}
h(r)\approx 1-\frac{ 2^{1-n}}{(n+2)(n+3)}\left(\frac{1}{M_\star \sqrt{\pi}}\right)^{n+1} \left(\frac{M}{M_\star}\right)
\left(\frac{1}{\sqrt{\theta}}\right)^{n+3}\ r^2.
\end{equation}
This is the signature of the regularity of the manifold at short scales. The de Sitter core is nothing but an effective geometry which accounts for the mean value of quantum gravity fluctuations and prevents the energy profile from collapsing into a Dirac delta, by means of a locally repulsive gravitational effect.
The gamma function (\ref{eq:gamma}) provides the smooth transition between the classical geometry at large radii and the effective quantum geometry at small radii.

\begin{figure}[h]
\begin{center}
\includegraphics[width=10cm]{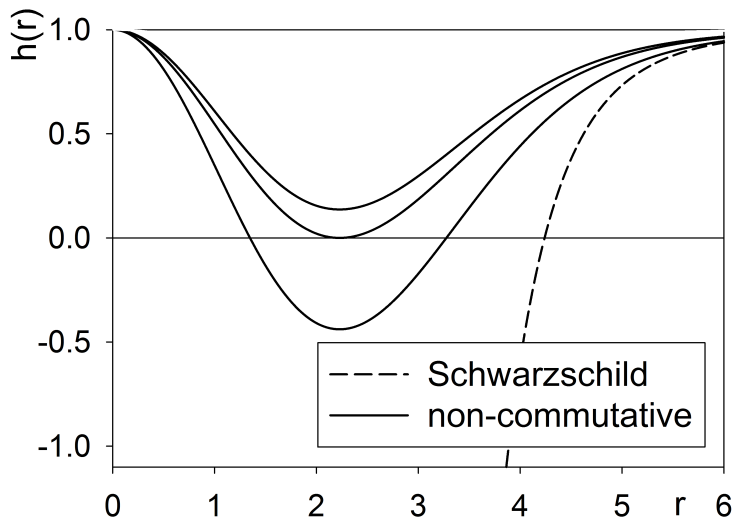}
\end{center}
\caption{Metric function $h(r)$ (\ref{eq:hr}) for various eleven-dimensional NCBH solutions (solid curves), illustrating the possible horizon structures. The dashed curve shows the same function for a higher-dimensional Schwarzschild black hole for comparison.}
\label{fig:metrics}
\end{figure}

Further features of the line element (\ref{eq:metric}) emerge by studying the horizon equation $h(r_h)=0$, a parametric equation depending on the mass parameter $M$ which is the integral of the energy density:
\begin{equation}
M=\frac{2\pi^{\frac{n+3}{2}}}{\Gamma\left(\frac{n+3}{2}\right)}\int_0^\infty dr\ r^{n+2} \  T_0^{\ 0}(r).
\end{equation}
There exists a threshold value $M_0$ for $M$ which lets us distinguish three cases:
\begin{enumerate}
\item for $M>M_0$ there exist two horizons, an inner Cauchy horizon $r_-$ and an outer event horizon $r_h$;
\item for $M<M_0$ there is no solution for $h(r_h)=0$ and no horizon occurs;
\item for $M=M_0$ the two horizons coalesce into a single degenerate event horizon $r_0$.
\end{enumerate}
These three possibilities are illustrated for eleven-dimensional black holes in figure~\ref{fig:metrics}.
The value of $M_0$ depends on $n$ and can be determined by numerical estimates \cite{Riz06,Nic09}.
The existence of an inner Cauchy horizon for $M>M_{0}$ opens the potential problem of the classical instability of the solution.  This is a feature that appears also in the case of LQBHs \cite{BMM11a,BMM11b} and has been investigated for NCBHs with controversial results \cite{BaN10,BrM11}. Even if a Cauchy horizon is certainly a surface of infinite blue shift where classically unbounded curvatures might develop, at a quantum level one may think that the same mechanism used to cure the curvature singularity might be invoked to tame divergent frequency modes in the vicinity of $r_-$. In any case, we can for now circumvent this problem, as in the case of classical Reissner-Nordstr\"{o}m or Kerr geometries, by saying that the potential instability would not become manifest within typical evaporation time scales, which have been proven to be extremely short \cite{CaN08}.

The no-horizon case corresponds to a manifold which is regular everywhere, an additional gravitational object, within a plethora of non-perturbative gravitational objects, that might be produced in super-Planckian collisions \cite{Cav03}.  In this class of no-horizon objects we have to consider also the case of spherically symmetric solutions that can be obtained by  flipping the sign of the radial coordinate $r\to -r$. Since the space-time is locally flat at the origin, the solution obtained by the $r\to - r$ map turns out to be geodesically complete. Therefore negative $r$ solutions are not merely analytic continuations of positive $r$ space-times, but genuinely new geometries \cite{MaN11}. The parity of the gamma function in (\ref{eq:hr}) implies that only for even $n$  we find distinct geometries by this procedure, which can be considered as geometries with positive $r$ and negative mass parameter $M$ (for more details about these geometries see \cite{Man97}). Finally, the last case, $M=M_0$, can be fully understood by studying the thermodynamic properties of the solutions since it is intimately related to the final configuration of the black hole at the end of the evaporation.

\begin{figure}[h]
\begin{center}
\includegraphics[width=12cm]{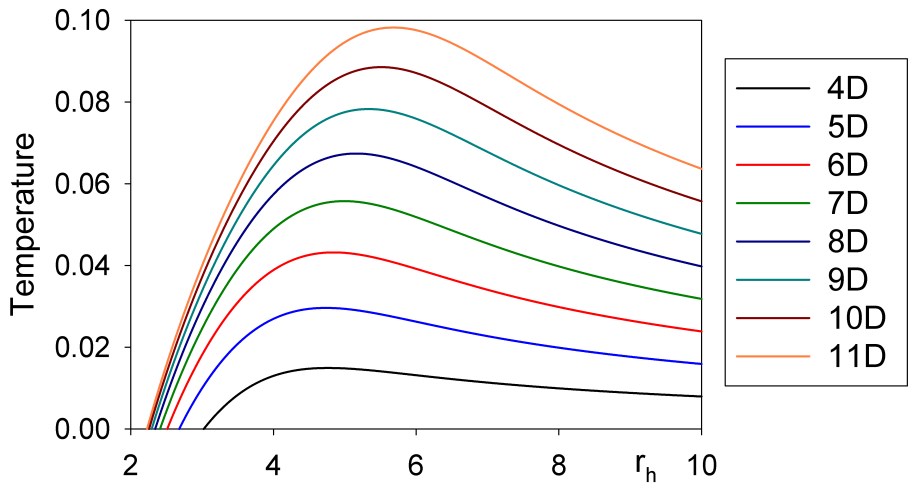}
\end{center}
\caption{The temperature $T$ (\ref{eq:temperature}) of NCBHs as a function of event horizon radius $r_{h}$, for various values of $n$, the number of extra dimensions, in units in which $M_{\star}={\sqrt {\theta }}=1$.
For $n>0$ this corresponds to energies in the TeV scale, while for $n=0$ energies are measured in units of $10^{16}$ TeV.
The different units are used for the $n=0$ case to facilitate comparison with the $n>0$ cases.
The temperatures increase with increasing $n$ for $n>0$, while the cooling phase leads to a smaller remnant for higher $n$.}
\label{fig:temp}
\end{figure}

The black hole temperature is given by
\begin{equation}
T = \frac{n+1}{4\pi r_{h}}\left[1-\frac{2}{n+1}\left(\frac{r_{h}}{2\sqrt{\theta}}\right)^{n+3}
\frac{e^{-r_{h}^2/4\theta}}{\gamma\left(\frac{n+3}{2}, \frac{r_{h}^2}{4\theta}\right)}\right].
\label{eq:temperature}
\end{equation}
We see that at large radii we recover the usual result $T\sim (n+1)/4\pi r_{h}$. However, at $r\sim\sqrt{\theta}$, quantum gravity corrections start to be dominant. As a consequence, in place of the usual divergent behavior for the temperature at small radii, there is a value at which the temperature vanishes. If we consider the internal energy of the system, by defining $M\equiv U(r_{h})$ as an implicit function of $r_{h}$ through the horizon equation $h(r_{h})=0$, we can show that it admits a minimum $M_0=U(r_0)$
\begin{equation}
\frac{dU(r_{h})}{dr_{h}}=
\frac{1}{8}(n+1)(n+2)\pi^{\frac{n+1}{2}}\frac{r_{h}^n M_\star^{n+2}}{\gamma\left(\frac{n+3}{2}, \frac{r_{h}^2}{4\theta}\right)} \left[1-\frac{2}{n+1}\left(\frac{r_{h}}{2\sqrt{\theta}}\right)^{n+3}
\frac{e^{-r_{h}^2/4\theta}}{\gamma\left(\frac{n+3}{2}, \frac{r_{h}^2}{4\theta}\right)}\right]
\end{equation}
for the same value of $r_0$ at which the temperature vanishes (for more details see \cite{Nic10}).
This implies that the extremal black hole case $M=M_0$ is actually a zero temperature configuration.
As the temperature is asymptotically vanishing, there should be a maximum temperature for some $r_{h}>r_0$, a fact that will have implications for the computation of the Hawking emission.
In conclusion, the temperature follows the usual curve at large radii, but as the black hole shrinks towards distances comparable with $\sqrt{\theta}$, it reaches a maximum temperature before cooling down towards a zero temperature extremal black hole remnant configuration (see figure~\ref{fig:temp}).

\begin{figure}
\begin{center}
\includegraphics[width=12cm]{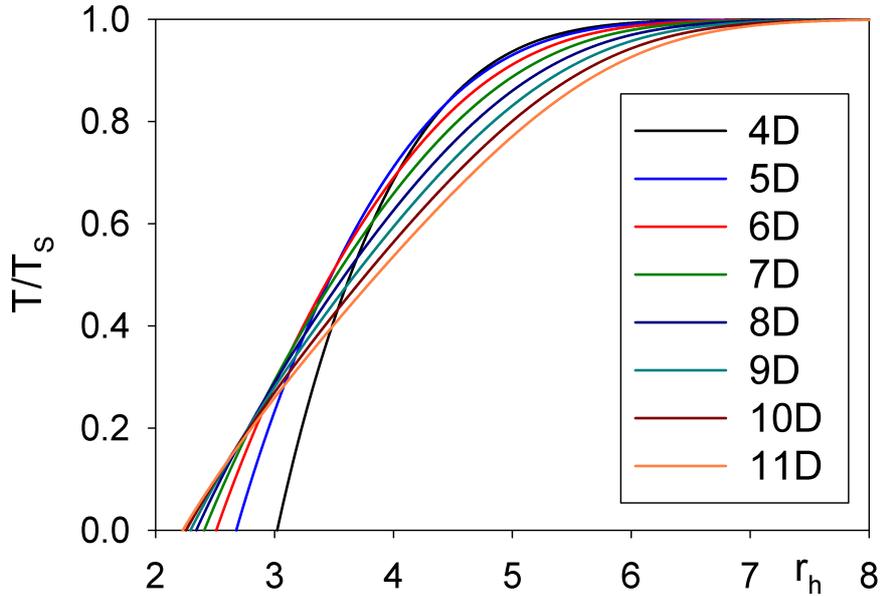}
\end{center}
\caption{The ratio $T /T_S$ as a function of $r_{h}$, in units in which
$M_{\star}={\sqrt {\theta }}=1$, for various values of $n$. Here $T$ is the temperature of an NCBH given by (\ref{eq:temperature}), and $T_{S}$ is the temperature of a classical Schwarzschild black hole having the same mass.
As $n$ increases, the value of $r_{h}$ for which the temperature is zero decreases, and the ratio $T/T_{S}$ for large $r_{h}$ also
decreases.
}
\label{fig:tempratio}
\end{figure}

At the maximum temperature, the system undergoes a phase transition from a locally unstable configuration with negative heat capacity, $C<0$ at large radii to a locally stable configuration at small radii with  $C>0$. The thermodynamic stability in the final phase of the evaporation is a feature that appears also in LQBHs and in ASGBHs. It has been argued that this is a general property of quantum gravity \cite{HuM09}. As a result, our analysis of Hawking emission could capture general features of the evaporation beyond the present case of NCBHs.
From figure~\ref{fig:tempratio} we see that quantum gravity effects become important in a region within $\sim 6\sqrt{\theta}$ from the origin, but are negligible for larger black holes. It is also clear from figure~\ref{fig:tempratio} that the temperature of an NCBH is considerably lower than that of a Schwarzschild black hole having the same mass. This will turn out to be the most important feature of the Hawking emission of NCBHs compared with higher-dimensional Schwarzschild black holes.

\begin{table}[h]
\begin{center}
\resizebox{\textwidth}{!}{
 \begin{tabular}{|c||c|c|c|c|c|c|c|c|}
\hline
 &  $n=0$  & $n=1$ & $n=2$ & $n=3$ & $n=4$ & $n=5$ & $n=6$ & $n=7$
\\
\hline
\hline
 $M_0$ (TeV)
&  $1.90 \times 10^{16}  $  &  $15.8$  & $102$  & $581$ &  $3.02\times 10^3$
 & $1.48\times 10^4$  &  $6.91\times 10^4$  &$3.13\times 10^5$\\
\hline
$r_0$ ($10^{-4}$ fm)   &  $3.02 \times 10^{-16}$   & $2.68$  & $2.51$  & $2.41$  & $2.34$  &
$2.29$  & $2.26$  & $2.23$ \\
\hline
\end{tabular} }
\end{center}
\caption{Minimum masses and minimum radii of NCBHs for different $n$. For $n=0$ the units are $M_\star\sim\sqrt{\theta}^{-1}\sim 10^{16} \ \mathrm{TeV}$.  For $n\neq 0$ the units are $M_\star\sim\sqrt{\theta}^{-1}\sim 1 \ \mathrm{TeV}$. }
 \label{tab:masses}
\end{table}

In view of particle physics experiments, we can now display potential values of the remnant size $r_0$ and mass $M_0$. In table~\ref{tab:masses} we see that the minimum mass to have black holes increases with $n$, while the corresponding radius slightly decreases. According to the latest experimental constraints, the tightest and most pessimistic estimate for the size of extra dimensions comes from the on-shell production of gravitons and sets $R\lesssim  10^{-12}\ \mathrm{m}$ \cite{Aad:2011xw,Chatrchyan:2011nd}. This limit requires $n\geq 3$ to have $M_\star$ at the terascale. This would imply that the NCBHs are too heavy to be produced at the LHC, at least as long as one assumes $\sqrt{\theta}\sim 10^{-4} \ \mathrm{fermi}$\footnote{In the present paper we have used the notation of Myers and Perry for the fundamental mass $M_\star$ \cite{MyP86}, which has also been adopted in \cite{Cav03,Kan04}. However, alternative definitions of the fundamental mass $M_\star$ have appeared in the literature. These lead to different values for the NCBH minimum masses. Our results are consistent with all previous findings. For instance, Rizzo obtained minimum masses that correspond to $8\pi M_0$, where $M_0$ is given in table \ref{tab:masses} \cite{Riz06}.  Gingrich's minimum masses correspond to $4(2\pi)^{1-n}M_0$ \cite{Gin10}, while the notation of Spallucci and coworkers leads to masses $\frac{4\Gamma(\frac{n+3}{2})}{(n+2)} \pi^{-\frac{n+1}{2}} M_0$ \cite{SSN09,Nic09}. Despite these different definitions, the conclusion is unique in all cases: if $M_\star\sqrt{\theta}=1$ minimum masses are not in the energy range accessible at the LHC for $n\geq 2$. }.  Alternatively, the possibility of a smaller minimal length has been considered, that is, $M_\star\sqrt{\theta}< 1$,
in order to get into the LHC-accessible black hole mass region \cite{Riz06}.
This possibility is based on the fact that in general we ignore the exact nature of the relation between $\sqrt{\theta}$ and the mass scale $\Lambda_{\mathrm{NC}}$ associated with the appearance of non-commutative effects. In other words, we cannot say anything more than $\sqrt{\theta}\propto 1/\Lambda_{\mathrm{NC}}$ and $\Lambda_{\mathrm{NC}}\sim M_\star$. As a consequence, we can proceed by setting the value of $M_\star\sqrt{\theta}$ in order to have the minimum mass in the range $1 \ \mathrm{TeV}\lesssim M_0\lesssim 10 \ \mathrm{TeV}$.
From $M=U(r_{h})$, we see that $M\sim r_{h}^{n+1} M_\star^{n+2}$. Once the radius $r_{h}$ is expressed in $\sqrt{\theta}$ units we find that
$M\propto (M_\star\sqrt{\theta})^{n+1} M_\star$. Thus we conclude that for $M_\star\sqrt{\theta}\approx 0.27$, the NCBH masses would be accessible to current particle physics experiments for all $n>0$. In addition, these threshold masses are compatible with recent limits established by experimental observations at the CMS detector \cite{CMS11}.

\begin{table}
\begin{center}
\resizebox{\textwidth}{!}{
\begin{tabular}{|c||c|c|c|c|c|c|c|c|}
\hline
&  $n=0$ & $n=1$ & $n=2$ & $n=3$ & $n=4$ & $n=5$ & $n=6$ & $n=7$
\\
\hline
\hline
 $T_{\mathrm{max}}$ (GeV)
&  $15\times 10^{16}$  & 30  & 43  & 56  & 67  & 78  & 89  & 98 \\
\hline $T_{\mathrm{max}}$ ($10^{15} K$) &  $0.18\times 10^{16}$  & 0.35  & 0.50  & 0.65  & 0.78  & 0.91  & 1.0
& 1.1
\\
\hline
\end{tabular} }
\end{center}
\caption{NCBHs maximum temperatures for different values of $n$ with units in which $M_\star\sqrt{\theta}=1$.}
\label{tab:temp}
\end{table}

However, having $M_\star\sqrt{\theta}<1$ has further repercussions on other parameters characterizing the physics of NCBHs. A smaller length scale not only gives smaller black hole masses but also smaller horizon radii and therefore higher temperatures. As a result, the quantum back-reaction which we claimed to be negligible could turn out to be relevant for such a choice of $\sqrt{\theta}$.
\begin{table}[h]
\begin{center}
\resizebox{\textwidth}{!}{
 \begin{tabular}{|c||c|c|c|c|c|c|c|c|}
\hline
&  $n=0$ & $n=1$ & $n=2$ & $n=3$ & $n=4$ & $n=5$ & $n=6$ & $n=7$
\\
\hline
\hline
  $ T/M  $
&  $< 1\times 10^{-2} $  & $< 2 \times 10^{-3} $  & $< 4\times 10^{-4} $  & $< 1\times 10^{-4} $  & $ < 2 \times 10^{-5} $  & $<
5 \times 10^{-6} $  & $< 1 \times 10^{-6}$  & $< 3\times 10^{-7}$ \\
\hline
\end{tabular} }
\end{center}
\caption{Quantum back reaction estimates for different $n$ with units in which $M_\star\sqrt{\theta}=1$.}
 \label{tab:back}
\end{table}
In table~\ref{tab:temp} we have an estimate of the maximum temperatures in the case $M_\star\sqrt{\theta}=1$. We see that the temperature can be at most $\sim 100\ \mathrm{GeV}$ for $n>0$. Consequently, the back reaction is negligible, since the ratio $T/M$ is always quite small (see table~\ref{tab:back}).
For $M_\star\sqrt{\theta}\approx0.27$ maximum temperatures are in the range $210\ \mathrm{GeV}\lesssim T_{\mathrm{max}}\lesssim 360\ \mathrm{GeV}$ for $n\geq 3$.
However the ratio $T/M$ will never exceed $T_{\mathrm{max}}/M_0$ for all $n$. Since the latter ratio goes like $T_{\mathrm{max}}/M_0\propto ( M_\star\sqrt{\theta})^{-(n+2)}$, we find that $T/M< T_{\mathrm{max}}/M_0\leq 0.07$ for $n\geq 3$. Thus we conclude that the back reaction can be still considered to be negligible in this regime of parameters.

Another potentially serious repercussion of having $M_\star\sqrt{\theta}<1$ is a decreased minimum NCBH production cross-section $\sigma_{\mathrm{NCBH}}\sim \pi r_0^2$.  A smaller length scale gives smaller remnant radii $r_0\sim (M_\star\sqrt{\theta})\ 10^{-4} \ \mathrm{fermi}$ and therefore a quadratically smaller cross-section. However, even for $M_\star\sqrt{\theta}=0.27$ we still find a promising value for the cross-section, namely, $\sigma\sim 10^{-38}\ \mathrm{m}^2\sim 100\ \mathrm{pb}$. Given the latest peak LHC luminosity $L\sim 3.2\times 10^{37}\  \mathrm{m}^{-2}\ \mathrm{s}^{-1}$ \cite{CERN1}, roughly a black hole every three seconds would be produced at the CERN laboratories, an astonishing number which does not differ significantly from that expected for conventional black hole metrics. Such a plentiful production of black holes might seem like a speculative prediction to a skeptical reader. However,  for our primary goal  in this paper (which is to study the differences in Hawking emission from NCBHs compared to classical Schwarzschild black holes), the condition $M_\star\sqrt{\theta}<1$ is irrelevant. For the rest of this paper we will work with the usual condition $M_\star\sqrt{\theta}=1$. This implies that our phenomenological predictions will have particular consequences for the physics of cosmic ray showers. Here the energy available can reach $\sim 10^{8} \ \mathrm{TeV}$, definitely much higher than that needed to produce NCBHs \cite{Cav03}.

\section{Hawking emission}
\label{sec:Hawkrad}

Hawking radiation from higher-dimensional black holes has been widely researched, see for example \cite{Cav03,Kan04,Hos05,Web05,CaS06,Win07,BlN10,Kanti:2008eq,Kanti:2009sz} for some reviews.
As well as being of intrinsic interest, a detailed quantitative understanding of the Hawking emission is essential for accurate simulations of mini black hole events at the LHC \cite{Frost:2009cf,Dai:2007ki}.
For spherically symmetric, higher-dimensional Schwarzschild black holes the Hawking radiation both on the brane and in the bulk has been extensively studied (some references are \cite{Cardoso:2005mh,Cardoso:2005vb,Cornell:2005ux,Creek:2006ia,Harris:2003eg,
Kanti:2002nr,Kanti:2002ge,Hod:2011zzb}),
including the graviton emission.
More recently, the emission from rotating higher-dimensional black holes has received attention (an incomplete list of references is
\cite{CKW06,CDK07,CDK08,Casals:2009st,Creek:2007pw,Creek:2007tw,
Creek:2007sy,Duffy:2005ns,Harris:2005jx,Flachi:2008yb,Frolov:2002as,Frolov:2002xf,Ida:2002ez,
Ida:2005ax,Ida:2006tf}).
The emission of massless particles of spin-zero, spin-one-half and spin-one on the brane, and spin-zero in the bulk, has been computed in detail and implemented in simulations of black hole events at the LHC \cite{Frost:2009cf,Dai:2007ki}.
However, only partial results are available for graviton emission
\cite{Kanti:2009sn,Doukas:2009cx}.
The most recent work on Hawking radiation from higher-dimensional black holes has focussed on the emission of particles of mass \cite{Kanti:2010mk,Rogatko:2009jp,Sampaio:2009tp,Sampaio:2009ra} or charge \cite{Sampaio:2009tp,Sampaio:2009ra}, or studying more complicated black hole geometries.  Of the latter, we mention only Gauss-Bonnet black holes
\cite{Grain:2005my,Konoplya:2010vz} which have a lower temperature compared with the usual Schwarzschild black holes, leading to a longer black hole lifetime \cite{Konoplya:2010vz}.

In this paper, we study the Hawking radiation of scalar fields from the black holes
(\ref{eq:metric}), both on the brane and in the bulk.  We focus on a scalar field because this is the simplest case and we anticipate that it will display many of the physical features of the emission common to all particle species.
Of course, the emission of particles of higher spin is important for phenomenology and we plan to return to this in a future publication.

For the moment therefore, we restrict our attention to a massless, minimally coupled scalar field satisfying the Klein-Gordon equation
\begin{equation}
\nabla _{\mu }\nabla ^{\mu } \Phi = 0 ,
\label{eq:KG}
\end{equation}
for comparison with previous results on the emission from Schwarzschild black holes \cite{Harris:2003eg}.
Since the black hole is non-rotating, and the scalar field uncharged, we are interested in the fluxes of particles and energy.
These fluxes are computed as expectation values of particle number operator and the component $T_{rt}$ of the stress-energy tensor respectively, the expectation
values being found in the Unruh vacuum \cite{Unruh:1976db} which models an evaporating black hole.
Fortunately we can compute these expectation values without recourse to curved-space
renormalization (see for example \cite{CDK08}).

We recall here that the expectation value of the stess-energy tensor for a quantized scalar field in a general space-time of arbitrary dimension is given as the limit \cite{DeF08}
\begin{equation}
\left<\psi\left|T_{\mu\nu}(x)\right|\psi\right>=\lim_{x\to x^\prime} \ \mathbb{D}_{\mu\nu}
(x,x^\prime)\ [-i G_F(x,x^\prime )]
\end{equation}
where $G_F(x,x^\prime )$ is the Feynman propagator
\begin{equation}
G_F(x,x^\prime )=i\left<\psi\left|T \Phi(x)\Phi(x^\prime)\right|\psi\right>
\end{equation}
(here $T$ denotes time ordering), $\left.\left.\right|\psi\right>$ is a normalized quantum state of Hadamard type and $\mathbb{D}_{\mu\nu} (x,x^\prime)$ is a differential operator given, for a massless, minimally coupled scalar field, by
\begin{equation}
\mathbb{D}_{\mu\nu}= g_\nu^{\ \nu^\prime}\nabla_\mu
\nabla_{\nu^\prime} - \frac {1}{2} g_{\mu\nu}g^{\rho\sigma^\prime}\nabla_\rho
\nabla_{\sigma^\prime}
\end{equation}
where $g_{\mu\nu^\prime}$ is the bivector of parallel transport from $x$ to $x^\prime$.
The Feynman propagator is, by definition, a solution of
\begin{equation}
\nabla _{\mu }\nabla ^{\mu } G_F(x,x^\prime )=-[-g(x)^{-1/2}]\ \delta^{D}(x-x^\prime )
\label{eq:green}
\end{equation}
where $D$ is the number of space-time dimensions. The issue is now to consider short-scale modifications not only of the gravity sector but of the matter sector too. In other words, we consider the possibility that the field $\Phi$ is affected by the presence of a quantum-gravity-induced minimal length, i.e.~a natural ultra-violet cut-off. Since this is the subject of much research, we consider only the results given in \cite{SSN06} as a preliminary step and we reserve the analysis of alternative modifications for forthcoming contributions.

The introduction of space-time fluctuations in quantum field theory can be achieved by considering a modified form of the Green function equation (\ref{eq:green}). By analogy with what we have seen on the gravity side, we model all the relevant modifications by a non-standard source term in the Green function equation (\ref{eq:green}). For mathematical convenience, we temporarily switch to Euclidean signature and we find
\beq
\Delta G_E(x,x^\prime)=e^{\frac {1}{2}\theta \Delta}\frac{1}{\sqrt{g}} \ \delta^{D}(x-x^\prime ),
\eeq
where $\Delta = \nabla _{\mu } \nabla ^{\mu }$   and the non-local operator $e^{\frac {1}{2}\theta \Delta}$ smears out any point-like object.
We introduce the Euclidean Green function $G_0(x,x^\prime )$ corresponding to the usual case $\sqrt{\theta}=0$, and then one can obtain the following relation between $G_0(x,x^\prime )$ and $G_E(x,x^\prime )$, namely:
\begin{equation}
G_E(x,x^\prime )=e^{\frac {1}{2} \theta \Delta}\ G_0(x,x^\prime ) .
\end{equation}
As a consequence, if we want to compute the stress-energy tensor corresponding to $G_E(x,x^\prime )$, we need to consider terms emerging from the following non-trivial commutation relation
\begin{equation}
\left[e^{\frac {1}{2} \theta \Delta}, \mathbb{D}_{\mu\nu}\right]\neq 0.
\end{equation}
The above expression will depend on curvature terms. However, given the regularity of our background metric we make the (brutal) approximation of neglecting these contributions just to have a flavour of the possible repercussions for the Hawking emission of the presence of the non-local operator. As a result, we model the effects of an effective ultra-violet cut-off in the frequency $\omega $ of the emitted quanta, which modifies the expectation values of the stress-energy tensor
in the following simplified way \cite{CaN08}:
\begin{equation}
\left<T_{rt}\right>\propto
\sum _{{\mathrm {modes}}}
\frac { e^{-\frac {1}{2}\theta\omega ^{2}} \omega }{
\left[ \exp \left( \omega /T \right) -1 \right] }
\left( 2 \ell + 1 \right)
\left| {\mathcal {A}}_{\omega \ell } \right| ^{2},
\label{eq:damping}
\end{equation}
where $\ell $ is a quantum number labelling a scalar field mode, $T$ is the Hawking temperature and ${\mathcal {A}}_{\omega \ell }$ is a transmission coefficient which will be defined in the following subsections.
This expression is phenomenologically motivated by the fact that all frequencies higher than $1/\sqrt{\theta}$ become largely suppressed.
Setting $\theta = 0$ in (\ref{eq:damping}), we recover the standard Hawking flux.
We comment that the above approximation is reasonable: we
have neglected curvature corrections that in the worst case are of order $R\sim 1/\theta$, corresponding to sub-leading
disturbances of frequencies $\omega \sim \sqrt{R} \sim
1/\sqrt{\theta}$.

The particle flux is not computed directly from the Feynman Green's function, but we model the effects on the particle flux of the non-local operator described above in the same way as for expectation values of the stress-energy tensor, namely by inserting a damping factor $e^{-\frac {1}{2} \theta \omega ^{2}}$ in the flux to give
\begin{equation}
\frac {dN}{dt} \propto \sum _{{\mathrm {modes}}}
\frac {e^{-\frac {1}{2}\theta\omega ^{2}}}{
\left[ \exp \left( \omega /T \right) -1 \right] }
\left( 2\ell + 1 \right)
\left| {\mathcal {A}}_{\omega \ell } \right| ^{2},
\label{eq:dampedN}
\end{equation}
which reduces to the usual Hawking flux when $\theta = 0$.
In the following sections, we shall compare the usual Hawking emission quantities with those originating from the above stress-energy tensor (\ref{eq:damping}) and particle flux.
From now on, throughout this section we use units in which the length scale ${\sqrt {\theta }}$ and $M_{\star}$ are set equal to unity, so that energies are in units of TeV for $n>0$ and $10^{16}$ TeV for $n=0$.  We shall consider scalar field modes of frequency up to $\omega =1$ in these units (corresponding to frequencies up to $1/{\sqrt {\theta }}$), since frequencies above this value are suppressed in our model.

\subsection{Emission on the brane}
\label{sec:brane}

To compute the emission of brane-localized modes, we consider a four-dimensional ``slice'' of the higher-dimensional black hole (\ref{eq:metric}) obtained from fixing the co-ordinates $\vartheta _{i}$, $i>1$ to give the
resulting metric
\begin{equation}
ds^{2} = -h(r) \, dt^{2} + h(r)^{-1} \, dr^{2} + r^{2} \left( d\vartheta ^{2}
+\sin ^{2} \vartheta \, d\varphi ^{2} \right) ,
\label{eq:branemetric}
\end{equation}
where we have set $\vartheta_1\equiv\vartheta$.
We perform the usual frequency decomposition of the scalar field $\Phi$ and consider
field modes of the form
\begin{equation}
\Phi ^{\mathrm {brane}} _{\omega \ell m} \left( t,r, \vartheta ,\varphi \right)  = e^{-i\omega t} e^{im\varphi }
R_{\omega \ell }^{\mathrm {brane}}(r) Y_{\ell }^{m} (\vartheta ) ,
\end{equation}
where $\omega $ is the frequency of the mode, $m$ the azimuthal quantum number, and
$Y_{\ell }^{m} (\vartheta )$ is a scalar spherical harmonic.
The radial function $R_{\omega \ell m}^{\mathrm {brane}}(r)$ satisfies the equation
\begin{equation}
0 = \frac {d}{dr} \left[ r^{2} h(r) \frac {dR_{\omega \ell }^{\mathrm {brane}}}{dr} \right]
+ \left[ \frac {\omega ^{2}r^{2}}{h(r)} - \ell \left( \ell + 1 \right) \right]
R_{\omega \ell }^{\mathrm {brane}}.
\label{eq:4dradial}
\end{equation}
A suitable basis of linearly independent solutions of the radial equation (\ref{eq:4dradial}) is given by
the ``in'' and ``out'' modes:
\begin{eqnarray}
R_{\omega \ell }^{\mathrm {brane,in}} & = &
\left\{
\begin{array}{ll}
e^{-i\omega r_{*}} & \qquad \qquad r \rightarrow r_{h}
\\
r^{-1} \left[ A_{\omega \ell }^{\mathrm {brane,in}} e^{-i\omega r_{*}} +
A_{\omega \ell }^{\mathrm {brane,out}} e^{i\omega r_{*}} \right]
& \qquad \qquad r\rightarrow \infty
\end{array}
\right.
\label{eq:4din}
\\
R_{\omega \ell }^{\mathrm {brane,out}} & = &
\left\{
\begin{array}{ll}
B_{\omega \ell }^{\mathrm {brane,in}} e^{-i\omega r_{*}}
+ B_{\omega \ell }^{\mathrm {brane,out}} e^{i\omega r_{*}}
& \qquad \qquad r\rightarrow r_{h}
\\
r^{-1} e^{i\omega r_{*}} & \qquad \qquad r\rightarrow \infty
\end{array}
\right.
\label{eq:4dout}
\end{eqnarray}
where we have defined the ``tortoise'' co-ordinate $r_{*}$ by
\begin{equation}
\frac {dr_{*}}{dr} = \frac {1}{h(r)}.
\label{eq:tortoise}
\end{equation}

The conventional particle flux spectrum, the number of particles emitted per unit time and unit frequency, is
\begin{equation}
\frac {d^{2}N^{\mathrm{ brane}}}{dt\, d\omega } =
\frac {1}{2\pi  }
\frac {1}{\exp \left( \omega /T \right) -1}
\sum _{\ell =0}^{\infty } \left( 2\ell + 1 \right)
\left| {\mathcal {A}}_{\omega \ell }^{\mathrm {brane}} \right| ^{2} ,
\label{eq:braneparticle}
\end{equation}
and the standard power spectrum, the energy emitted per unit time and unit frequency, is
\begin{equation}
\frac {d^{2}E^{\mathrm {brane}}}{dt \, d\omega } =
\frac {1}{2\pi  }
\frac {\omega }{\exp \left( \omega /T \right) -1}
\sum _{\ell =0}^{\infty } \left( 2\ell + 1 \right)
\left| {\mathcal {A}}_{\omega \ell }^{\mathrm {brane}} \right| ^{2}  ,
\label{eq:branepower}
\end{equation}
where $T$ is the black hole temperature (\ref{eq:temperature}).
In (\ref{eq:braneparticle}--\ref{eq:branepower}), the quantity
$\left| {\mathcal {A}}_{\omega \ell }^{\mathrm {brane}}\right| ^{2}$ is the transmission coefficient for the scalar field mode.
If we consider a scalar wave which, near the event horizon, is out-going, the transmission coefficient is given by the proportion of the wave which tunnels through the gravitational potential surrounding the black hole and escapes to infinity.
We compute the transmission coefficients numerically from the ``in'' modes (\ref{eq:4din}) as follows:
\begin{equation}
\left| {\mathcal {A}}_{\omega \ell }^{\mathrm {brane}} \right| ^{2} =
1 - \frac {\left| A_{\omega \ell }^{\mathrm {brane,out}} \right| ^{2}}{
\left| A_{\omega \ell }^{\mathrm {brane,in}} \right| ^{2}}.
\label{eq:trans}
\end{equation}
As well as the usual particle and energy fluxes (\ref{eq:braneparticle}--\ref{eq:branepower}), we also study the fluxes discussed at the start of this section, where there is an additional damping term due to the non-commutativity (\ref{eq:damping}--\ref{eq:dampedN}) \cite{CaN08}:
\begin{eqnarray}
\frac {d^{2}N^{{\mathrm {brane,NC}}}}{dt \, d\omega } & = &
\frac {1}{2\pi }
\frac {e^{-\frac {1}{2}\omega ^{2}}}{
\left[ \exp \left( \omega /T \right) -1 \right] }
\sum _{\ell =0}^{\infty } \left( 2\ell + 1 \right)
\left| {\mathcal {A}}_{\omega \ell }^{\mathrm {brane}} \right| ^{2} ,
\label{eq:braneparticleNC}
\\
\frac {d^{2}E^{{\mathrm {brane,NC}}}}{dt \, d\omega } & = &
\frac {1}{2\pi }
\frac {\omega \, e^{-\frac {1}{2}\omega ^{2}}}{
\left[ \exp \left( \omega /T \right) -1 \right] }
\sum _{\ell =0}^{\infty } \left( 2\ell + 1 \right)
\left| {\mathcal {A}}_{\omega \ell }^{\mathrm {brane}} \right| ^{2}  .
\label{eq:branepowerNC}
\end{eqnarray}
We also consider the absorption cross-section
$\sigma ^{\mathrm {brane}}(\omega )$, which has the form
\begin{equation}
\sigma ^{\mathrm {brane}}(\omega )=
\frac {\pi }{\omega ^{2}}
\sum _{\ell =0}^{\infty }
\left( 2\ell +1 \right)
\left| {\mathcal {A}}_{\omega \ell }^{\mathrm {brane}} \right| ^{2}.
\label{eq:braneabsorption}
\end{equation}

\begin{figure}[hb]
\begin{center}
\includegraphics[width=9cm]{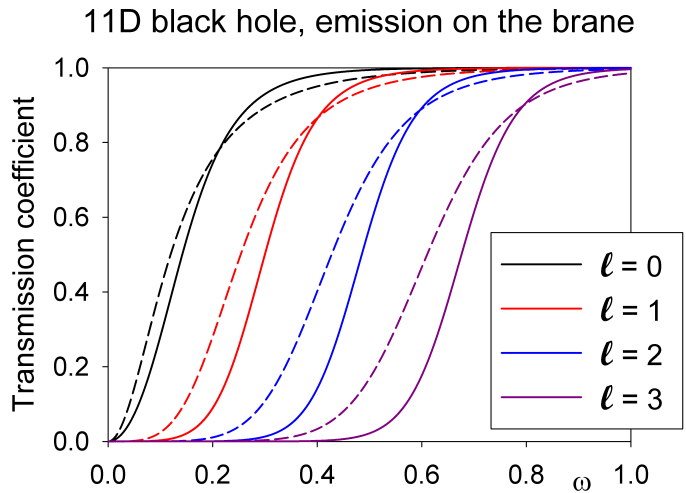}
\end{center}
\caption{Transmission coefficients (\ref{eq:trans}) for a scalar field on the brane as a function of frequency
$\omega $ for the first few modes. We consider an eleven-dimensional NCBH.   Solid lines are the transmission coefficients for the NCBH, while dotted lines denote the transmission coefficients for a Schwarzschild black hole with the same mass. The quantum number $\ell $ increases from $\ell =0$ to $\ell =3$ going from left to right. We use units in which $M_{\star}$ and $\theta $ are set equal to unity, so that energies are in TeV.}
\label{fig:trans}
\end{figure}

We begin the presentation of our numerical results by considering the transmission
coefficient (\ref{eq:trans}) (see figure~\ref{fig:trans} for the transmission coefficients for the first few modes for an eleven-dimensional black hole), and comparing them with those for a Schwarzschild black hole having the same mass.
For low frequencies, the transmission coefficient for the NCBH is smaller than that for the Schwarzschild black hole, but it converges more rapidly to unity as the frequency increases than in the Schwarzschild case.  These differences become more marked as the quantum number $\ell $ increases.

\begin{figure}[h]
\begin{center}
\includegraphics[width=11cm]{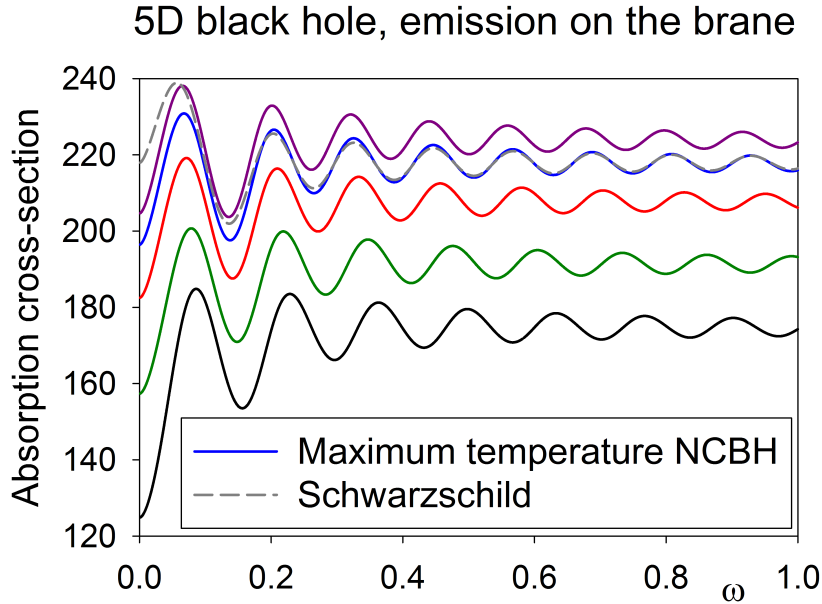}
\end{center}
\caption{Absorption cross-section (\ref{eq:braneabsorption}) for a scalar field on the brane as a function of frequency. Five-dimensional NCBHs with varying masses are considered, together with a
Schwarzschild black hole having the same mass as the NCBH with maximum temperature.
The dark grey dotted curve (having the largest low-frequency absorption cross-section) is for the Schwarzschild black hole, the other curves are for NCBHs, with the mass of the NCBH increasing as the value of the low-frequency absorption cross-section increases.  The blue curve (third from the top in the low-frequency limit) is the curve for the NCBH having maximum temperature.
We use units in which $M_{\star}$ and $\theta $ are set equal to unity, corresponding to energies measured in TeV.}
\label{fig:5dabsorptionbrane}
\end{figure}
\begin{figure}
\begin{center}
\includegraphics[width=12cm]{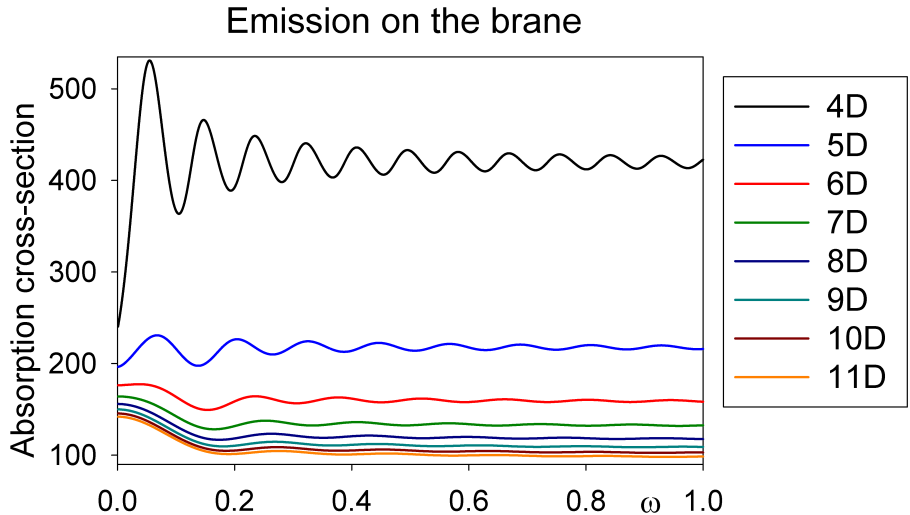}
\end{center}
\caption{Absorption cross-section (\ref{eq:braneabsorption}) for a scalar field on the brane as a function of frequency. NCBHs with maximum temperatures are considered for varying numbers of space-time dimensions. The curves, from top to bottom, are for $n=0$ up to $n=7$.  We use units in which $M_{\star}$ and $\theta $ are set equal to unity.  For $n>0$ this corresponds to energies measured in TeV, but for $n=0$ energies are measured in units of $10^{16}$ TeV.  The different units are used for the $n=0$ case to facilitate comparison with the $n>0$ cases.}
\label{fig:absorptionbrane}
\end{figure}

This behaviour of the transmission coefficients is reflected in the absorption cross-section (\ref{eq:braneabsorption}) as a function of frequency $\omega $,
see figures~\ref{fig:5dabsorptionbrane} and \ref{fig:absorptionbrane}.
The behaviour of the absorption cross-section on the brane for NCBHs is qualitatively similar to that observed for Schwarzschild black holes \cite{Harris:2003eg}.  As the frequency $\omega \rightarrow 0$, the absorption
cross-section tends to the area of the event horizon $4\pi r_{h}^{2}$,
as observed for brane emission from Schwarzschild black holes \cite{Harris:2003eg}.
For five-dimensional black holes (figure~\ref{fig:5dabsorptionbrane}), the area of the event horizon increases as the mass of the black hole increases, leading to the observed increase in the low-energy absorption cross-section. The difference in low-energy absorption cross-section between the NCBH with the maximum temperature and the Schwarzschild black hole with the same mass is due to the latter having a considerably larger event horizon area. For NCBHs with maximal temperature
(figure~\ref{fig:absorptionbrane}), the event horizon area decreases as the number of space-time dimensions increases, which gives the observed decrease in the low-energy absorption cross-section.

As $\omega $ increases, the absorption cross-section oscillates about its asymptotic high-energy value (although the magnitude of the oscillations decreases significantly as the number of space-time dimensions increases \cite{Hod:2011zzb}).  The high-frequency limiting value (also known as the geometric optics limit) of the total absorption cross-section corresponds to an effective absorbing area of radius $r_{c}$, where
$r_{c}^{2}=\sigma ^{\mathrm {brane}} (\omega \rightarrow \infty ) /\pi $.
From figure~\ref{fig:5dabsorptionbrane}, it can be seen that
$r_{c}$ increases as the mass of the five-dimensional black holes increases, and, furthermore, that $r_{c}$ is always greater than the event horizon radius $r_{h}$ for these black holes. The high-frequency absorption cross-sections for the NCBH with maximum temperature and the Schwarzschild black hole with the same mass are identical, indicating that the high-frequency absorption cross-section depends only on the mass of the black hole and not the detailed structure of the metric near the event horizon.
Looking at figure~\ref{fig:absorptionbrane}, we observe that for four- and five-dimensional black holes, the effective high-frequency absorption area is larger than the black hole event horizon, whilst for black holes in six and more dimensions, the effective high-frequency absorption area is smaller than the black hole event horizon area.

For the brane emission of scalar particles from an $\left( n + 4 \right)$-dimensional
black hole, it has been found that, for Schwarzschild black holes, $r_{c}$ is
related to the event horizon radius \cite{Emparan:2000rs}:
\begin{equation}
r_{c}= \left( \frac {n+3}{2} \right) ^{\frac {1}{\left( n + 1\right)}}
{\sqrt {\frac {n+3}{n+1}}} r_{h}.
\label{eq:rc}
\end{equation}
For the black holes in figures~\ref{fig:5dabsorptionbrane} and \ref{fig:absorptionbrane}, we find that $\sigma (\omega \rightarrow \infty ) =\pi r_{c}^{2}$ is a good approximation to the absorption cross-section when $\omega = 1$ if $r_{c}$ is given by (\ref{eq:rc}) with $r_{h}$ being the radius of a Schwarzschild black hole having the same mass as the NCBH.
In other words, as far as high-frequency modes are concerned, the NCBH is mimicking a Schwarzschild black hole with the same mass.

\begin{figure}
\begin{center}
\includegraphics[width=10cm]{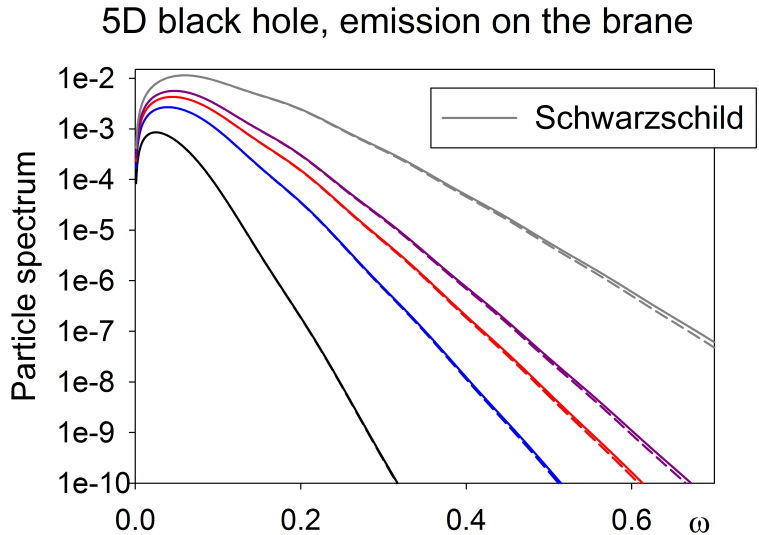}
\end{center}
\caption{Particle fluxes (\ref{eq:braneparticle}, \ref{eq:braneparticleNC}) for scalar field emission on the brane, as a function of frequency, for
five-dimensional NCBHs with varying masses and a Schwarzschild black hole having the same mass as the NCBH with maximum temperature.
Solid lines indicate the standard particle flux (\ref{eq:braneparticle}), and dotted lines the particle flux with additional damping due to non-commutative effects (\ref{eq:braneparticleNC}). The top curve (dark grey) is the flux for the Schwarzschild black hole, with the other curves being for NCBHs with increasing temperature from the bottom black curve to the top purple curve.  We use units in which $M_{\star}$ and $\theta $ are set equal to unity, corresponding to energies measured in TeV.}
\label{fig:5dbraneparticle}
\end{figure}
\begin{figure}
\begin{center}
\includegraphics[width=11cm]{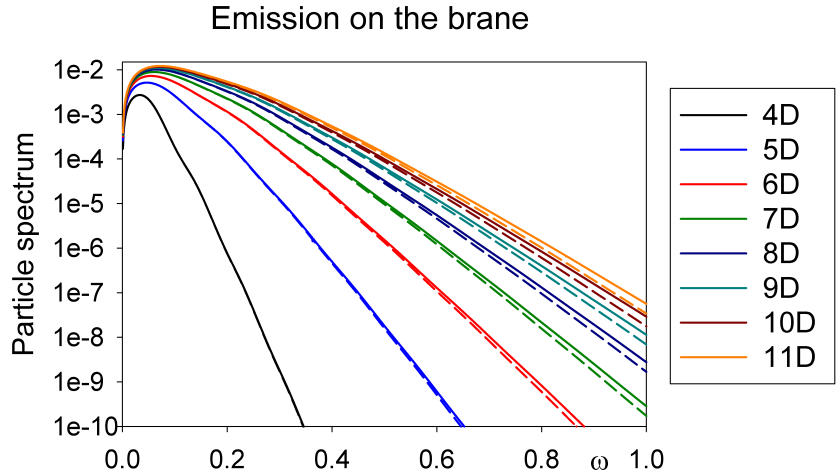}
\end{center}
\caption{
Particle fluxes (\ref{eq:braneparticle}, \ref{eq:braneparticleNC}) for scalar field emission on the brane, as a function of frequency, for NCBHs with maximum temperature and varying numbers of space-time dimensions.
The curves, from bottom to top, are for $n=0$ up to $n=7$.
Solid lines indicate the standard particle flux (\ref{eq:braneparticle}), and dotted lines the particle flux with additional damping due to non-commutative effects (\ref{eq:braneparticleNC}).  We use units in which $M_{\star}$ and $\theta $ are set equal to unity.  For $n>0$ this corresponds to energies measured in TeV, but for $n=0$ energies are measured in units of $10^{16}$ TeV.  The different units are used for the $n=0$ case to facilitate comparison with the $n>0$ cases.}
\label{fig:braneparticle}
\end{figure}

\begin{figure}
\begin{center}
\includegraphics[width=10cm]{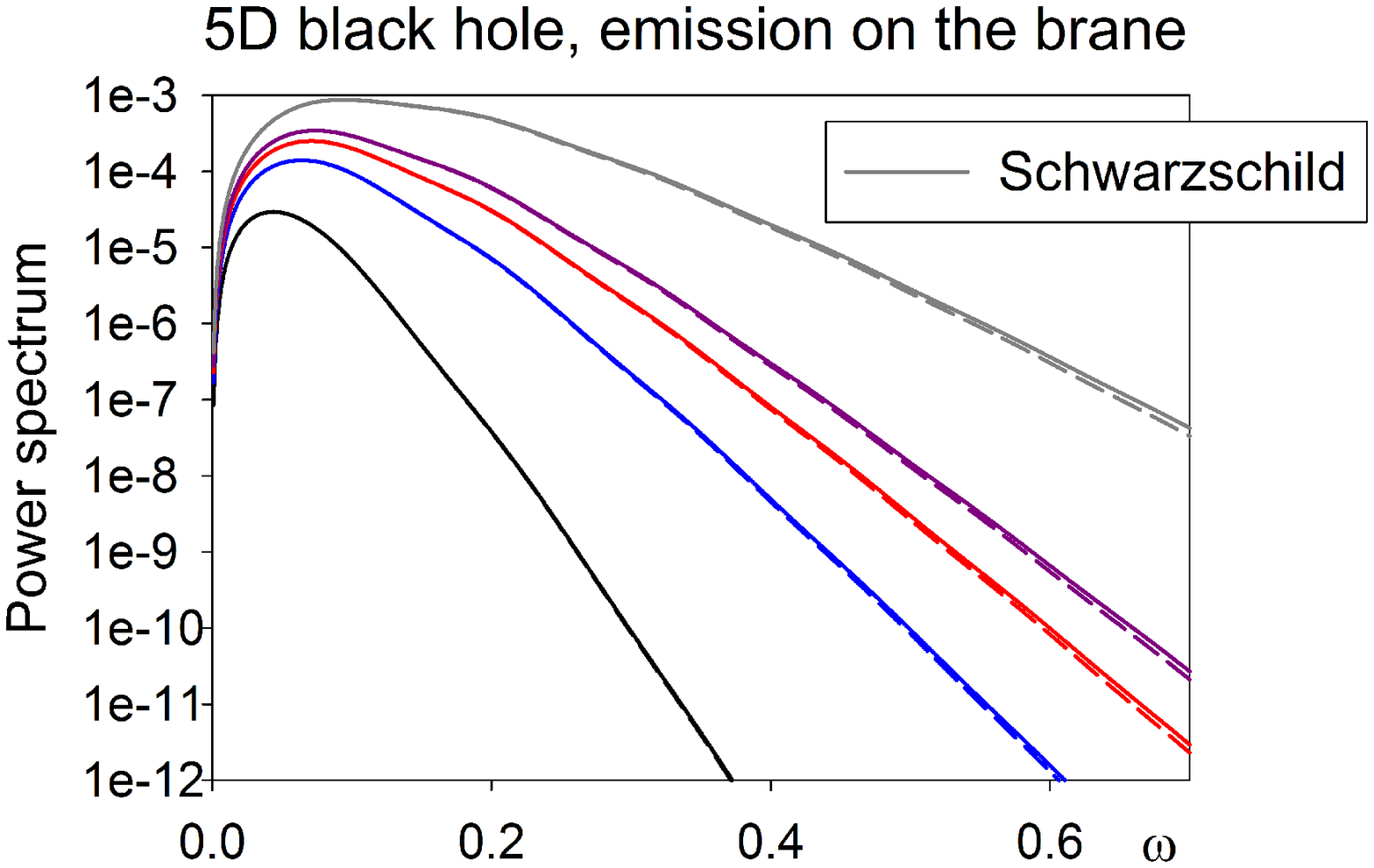}
\end{center}
\caption{Energy fluxes (\ref{eq:branepower}, \ref{eq:branepowerNC}) for scalar field emission on the brane, as a function of frequency. The same black holes are considered as in figure~\ref{fig:5dbraneparticle}.
As in figure~\ref{fig:5dbraneparticle}, solid lines correspond to the standard energy flux (\ref{eq:branepower}) and dotted lines the energy flux (\ref{eq:branepowerNC})
with additional damping due to non-commutative effects.
The top curve (dark grey) is the flux for the Schwarzschild black hole, with the other curves being for NCBHs with increasing temperature from the bottom black curve to the top purple curve.  We use units in which $M_{\star}$ and $\theta $ are set equal to unity, corresponding to energies measured in TeV.}
\label{fig:5dbranepower}
\end{figure}
\begin{figure}
\begin{center}
\includegraphics[width=11cm]{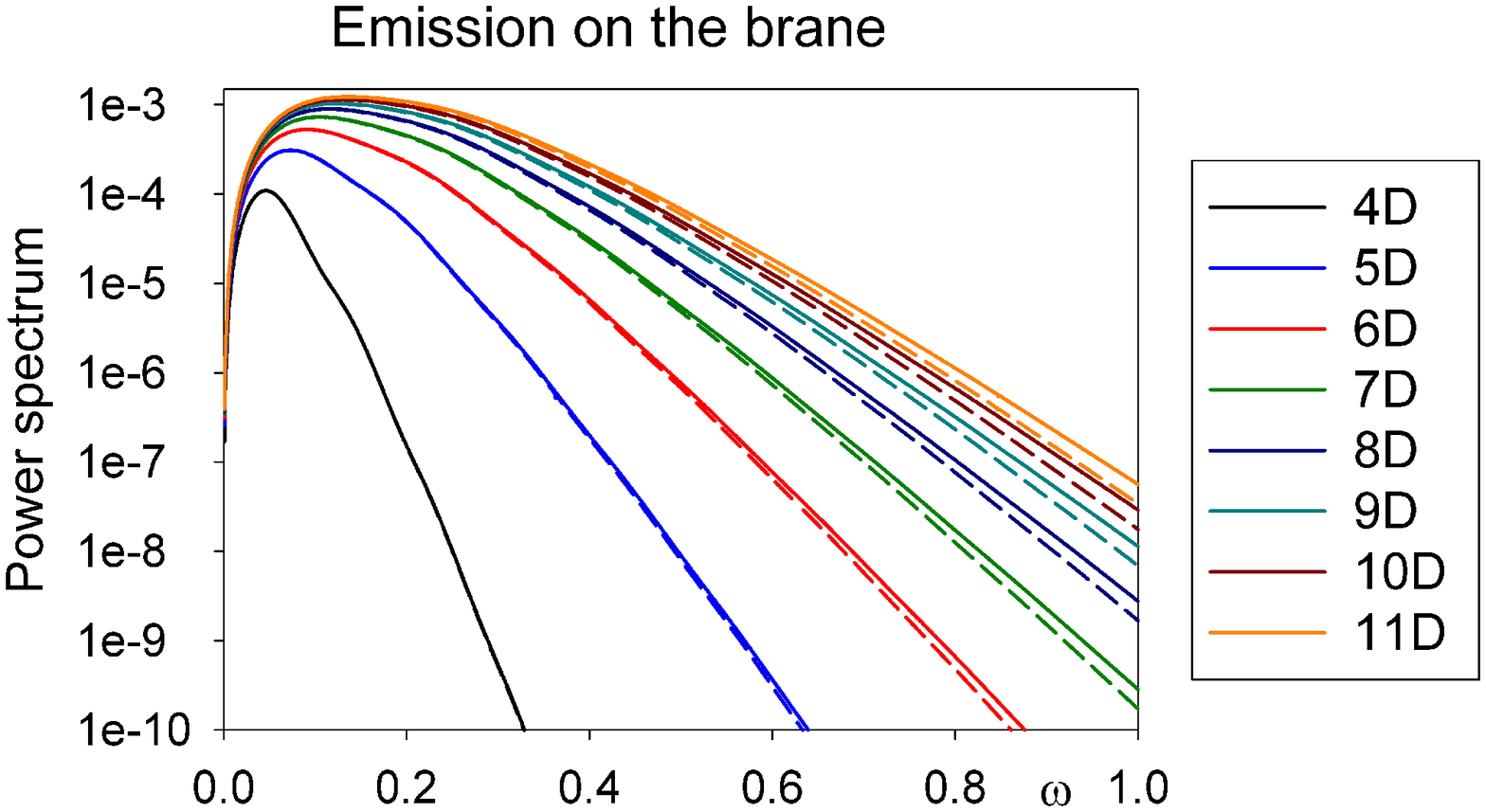}
\end{center}
\caption{Energy fluxes (\ref{eq:branepower}, \ref{eq:branepowerNC}) for scalar field emission on the brane, as a function of frequency. The same black holes are considered as in figure~\ref{fig:braneparticle}.
The curves, from bottom to top, are for $n=0$ up to $n=7$.
As in figure~\ref{fig:braneparticle}, solid lines correspond to the standard energy flux (\ref{eq:branepower}) and dotted lines the energy flux (\ref{eq:branepowerNC})
with additional damping due to non-commutative effects. We use units in which $M_{\star}$ and $\theta $ are set equal to unity.  For $n>0$ this corresponds to energies measured in TeV, but for $n=0$ energies are measured in units of $10^{16}$ TeV.  The different units are used for the $n=0$ case to facilitate comparison with the $n>0$ cases.}
\label{fig:branepower}
\end{figure}

Next we consider the scalar particle flux (\ref{eq:braneparticle}, \ref{eq:braneparticleNC}), see figures~\ref{fig:5dbraneparticle} and \ref{fig:braneparticle}, and scalar energy flux (\ref{eq:branepower}, \ref{eq:branepowerNC}), see figures~\ref{fig:5dbranepower} and \ref{fig:branepower}.
The results for particle and energy emission are very similar.  Looking first at the emission from five-dimensional NCBHs with varying masses (figures~\ref{fig:5dbraneparticle} and \ref{fig:5dbranepower}),
the flux is much smaller for NCBHs than from the Schwarzschild black hole having the same mass as the NCBH with maximum temperature. This is due to the considerably smaller temperature of the the NCBHs.  The peak of the emission in all cases comes from low-frequency modes, which probe more fully the nature of the black hole geometry near the horizon, for which there are differences in absorption cross-section (see figure~\ref{fig:absorptionbrane}) between NCBHs and Schwarzschild black holes.
However, the dominant effect across all frequency ranges is the much lower temperature of NCBHs compared with Schwarzschild black holes having the same mass.
For five-dimensional black holes, the additional damping due to the non-commutativity makes a negligible difference to the particle and energy flux at high frequency (recall that $\omega =1$ in our units corresponds to the extremely high frequency $\omega = \frac {1}{{\sqrt {\theta }}}$, where $\theta $ is the minimal length scale).

If we increase the number of space-time dimensions (figures~\ref{fig:braneparticle} and \ref{fig:branepower}), the maximum temperature of the NCBHs increases (see figure~\ref{fig:temp})  and we observe a corresponding increase in the scalar field emission. The peak of the particle and power spectra increase, and the emission remains significant for larger frequencies.  However, even for an 11-dimensional NCBH, the emission is still only of the same order of magnitude as a five-dimensional Schwarzschild black hole. The increase in brane emission as the number of space-time dimensions increases is much smaller than for Schwarzschild black holes \cite{Harris:2003eg}.
The other feature in figures~\ref{fig:braneparticle} and \ref{fig:branepower} is that the additional damping due to non-commutativity (\ref{eq:branepowerNC}) becomes more important as the number of space-time dimensions increases and the temperature of the black holes increases.

\subsection{Emission in the bulk}
\label{sec:bulk}

To study the bulk scalar modes, the Klein-Gordon equation (\ref{eq:KG}) must be solved on the full, higher-dimensional space-time (\ref{eq:metric}).  The scalar field modes
now take the form \cite{Harris:2003eg}
\begin{equation}
\Phi ^{{\mathrm {bulk}}}_{\omega \ell j}(t,r, \theta _{i}) =
e^{-i\omega t}R^{\mathrm {bulk}}_{\omega \ell }(r)
Y_{\ell }^{j}(\theta _{i}, \varphi ),
\end{equation}
where $Y_{\ell }^{j}(\theta _{i}, \varphi )$ is a scalar hyperspherical harmonic function of $\theta _{1}, \ldots , \theta _{n+1}, \varphi $ \cite{Muller}.
For each $\ell $ (which is the quantum number governing the constant arising in the separation of the Klein-Gordon equation), there are ${\mathcal {N}}_{\ell }$ hyperspherical harmonics, which we label by the index $j$.
The degeneracy factor ${\mathcal {N}}_{\ell }$ is
\begin{equation}
{\mathcal {N}}_{\ell }= \frac {\left( 2 \ell + n + 1\right) \,
\left( \ell + n \right) !}{
\ell ! \, \left( n + 1 \right) !},
\label{eq:hyperdegen}
\end{equation}
which reduces to the familiar $2\ell + 1$ when the number of extra dimensions, $n$,
is equal to zero.

The radial functions $R_{\omega \ell }^{\mathrm {bulk}}(r)$ now satisfy the equation
\begin{equation}
0= \frac {1}{r^{n}} \frac {d}{dr} \left[ h(r) r^{n+2}
\frac {dR_{\omega \ell }^{\mathrm {bulk}}}{dr} \right]
+ \left[ \frac {\omega ^{2}r^{2}}{h(r)} - \ell \left( \ell + n + 1 \right) \right]
R_{\omega \ell }^{\mathrm {bulk}}.
\end{equation}
The ``in'' and ``out'' radial functions (\ref{eq:4din}--\ref{eq:4dout}) are modified near infinity:
\begin{eqnarray}
R_{\omega \ell }^{\mathrm {bulk,in}} & = &
\left\{
\begin{array}{ll}
e^{-i\omega r_{*}} & \qquad \qquad r \rightarrow r_{h}
\\
r^{-1-\frac {n}{2}} \left[ A_{\omega \ell }^{\mathrm {bulk,in}} e^{-i\omega r_{*}} +
A_{\omega \ell }^{\mathrm {bulk,out}} e^{i\omega r_{*}} \right]
& \qquad \qquad r\rightarrow \infty
\end{array}
\right.
\label{eq:hdin}
\\
R_{\omega \ell }^{\mathrm {bulk,out}} & = &
\left\{
\begin{array}{ll}
B_{\omega \ell }^{\mathrm {bulk,in}} e^{-i\omega r_{*}}
+ B_{\omega \ell }^{\mathrm {bulk,out}} e^{i\omega r_{*}}
& \qquad \qquad r\rightarrow r_{h}
\\
r^{-1-\frac {n}{2}} e^{i\omega r_{*}} & \qquad \qquad r\rightarrow \infty
\end{array}
\right.
\label{eq:hdout}
\end{eqnarray}
where the ``tortoise'' co-ordinate $r_{*}$ is given in (\ref{eq:tortoise}).
The bulk particle and power spectra are most simply written as
\begin{eqnarray}
\frac {d^{2}N^{\mathrm {bulk}}}{dt \, d\omega }
& = & \frac {1}{2\pi }
\frac {1}{\exp \left( \omega /T \right) -1}
\sum _{\ell =0}^{\infty } {\cal {N}}_{\ell }
\left| {\mathcal {A}}_{\omega \ell }^{\mathrm {bulk}} \right| ^{2},
\label{eq:bulkparticle}
\\
\frac {d^{2}E^{\mathrm {bulk}}}{dt \, d\omega }
& = & \frac {1}{2\pi }
\frac {\omega }{\exp \left( \omega /T \right) -1}
\sum _{\ell =0}^{\infty } {\cal {N}}_{\ell }
\left| {\mathcal {A}}_{\omega \ell }^{\mathrm {bulk}} \right| ^{2},
\label{eq:bulkpower}
\end{eqnarray}
and we will also consider bulk particle and power spectra with additional damping terms due to non-commutativity effects:
\begin{eqnarray}
\frac {d^{2}N^{\mathrm {bulk,NC}}}{dt \, d\omega }
& = & \frac {1}{ 2\pi }
\frac {e^{-\frac {1}{2}\omega ^{2}}}{
\left[ \exp \left( \omega /T \right) -1 \right] }
\sum _{\ell =0}^{\infty } {\cal {N}}_{\ell }
\left| {\mathcal {A}}_{\omega \ell }^{\mathrm {bulk}} \right| ^{2},
\label{eq:bulkparticleNC}
\\
\frac {d^{2}E^{\mathrm {bulk,NC}}}{dt \, d\omega }
& = & \frac {1}{2\pi }
\frac {\omega \ e^{-\frac {1}{2}\omega ^{2}} }{
\left[ \exp \left( \omega /T \right) -1 \right]}
\sum _{\ell =0}^{\infty } {\cal {N}}_{\ell }
\left| {\mathcal {A}}_{\omega \ell }^{\mathrm {bulk}} \right| ^{2},
\label{eq:bulkpowerNC}
\end{eqnarray}
In the above equations (\ref{eq:bulkparticle}--\ref{eq:bulkpowerNC}), the bulk transmission coefficient
$\left| {\mathcal {A}}_{\omega \ell }^{\mathrm {bulk}} \right| ^{2}$
 appears, and this is computed in the same way as for the brane emission:
\begin{equation}
\left| {\mathcal {A}}_{\omega \ell }^{\mathrm {bulk}} \right| ^{2} =
1 - \frac {\left| A_{\omega \ell }^{\mathrm {bulk,out}} \right| ^{2}}{
\left| A_{\omega \ell }^{\mathrm {bulk,in}} \right| ^{2}}.
\end{equation}
We can also define a bulk absorption cross-section
$\sigma ^{\mathrm {bulk}}(\omega )$ in terms of the transmission
coefficient \cite{Harris:2003eg}:
\begin{equation}
\sigma ^{\mathrm {bulk}} (\omega ) =
\frac {2^{n}\pi ^{\frac {\left( n+1\right) }{2}}}{\omega ^{n+2}}
\left( n + 1 \right) \Gamma \left( \frac {n+1}{2} \right)
\sum _{\ell =0}^{\infty }
{\mathcal {N}}_{\ell } \left| {\mathcal {A}}_{\omega \ell }^{\mathrm {bulk}} \right| ^{2}.
\label{eq:bulkabsorption}
\end{equation}

\begin{figure}
\begin{center}
\includegraphics[width=10cm]{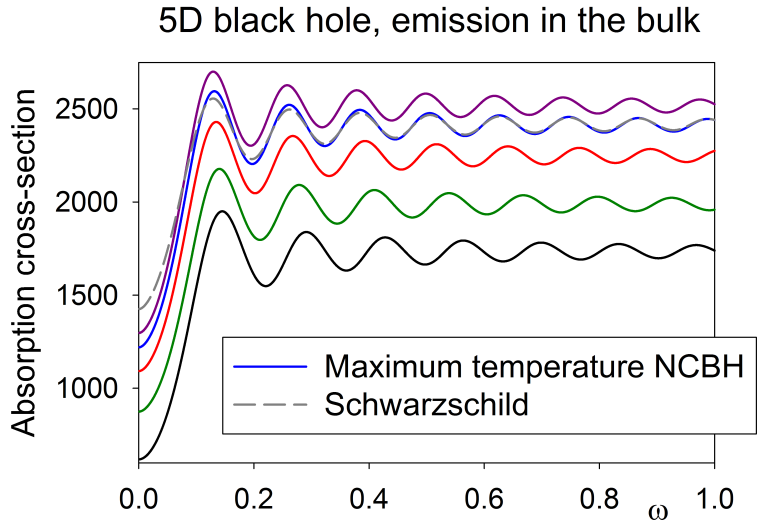}
\end{center}
\caption{Absorption cross-section (\ref{eq:bulkabsorption}) for a scalar field in the bulk as a function of frequency, for five-dimensional non-commutative black holes with varying masses, together with a
Schwarzschild black hole having the same mass as the non-commutative black hole with maximum temperature.
The dark grey curve (having the largest low-frequency absorption cross-section) is for the Schwarzschild black hole, the other curves are for NCBHs, with the mass of the NCBH increasing as the value of the low-frequency absorption cross-section increases.  The blue curve (third from the top in the low-frequency limit) is the curve for the NCBH having maximum temperature.
We use units in which $M_{\star}$ and $\theta $ are set equal to unity, corresponding to energies measured in TeV.}
\label{fig:5dabsorptionbulk}
\end{figure}
\begin{figure}
\begin{center}
\includegraphics[width=12cm]{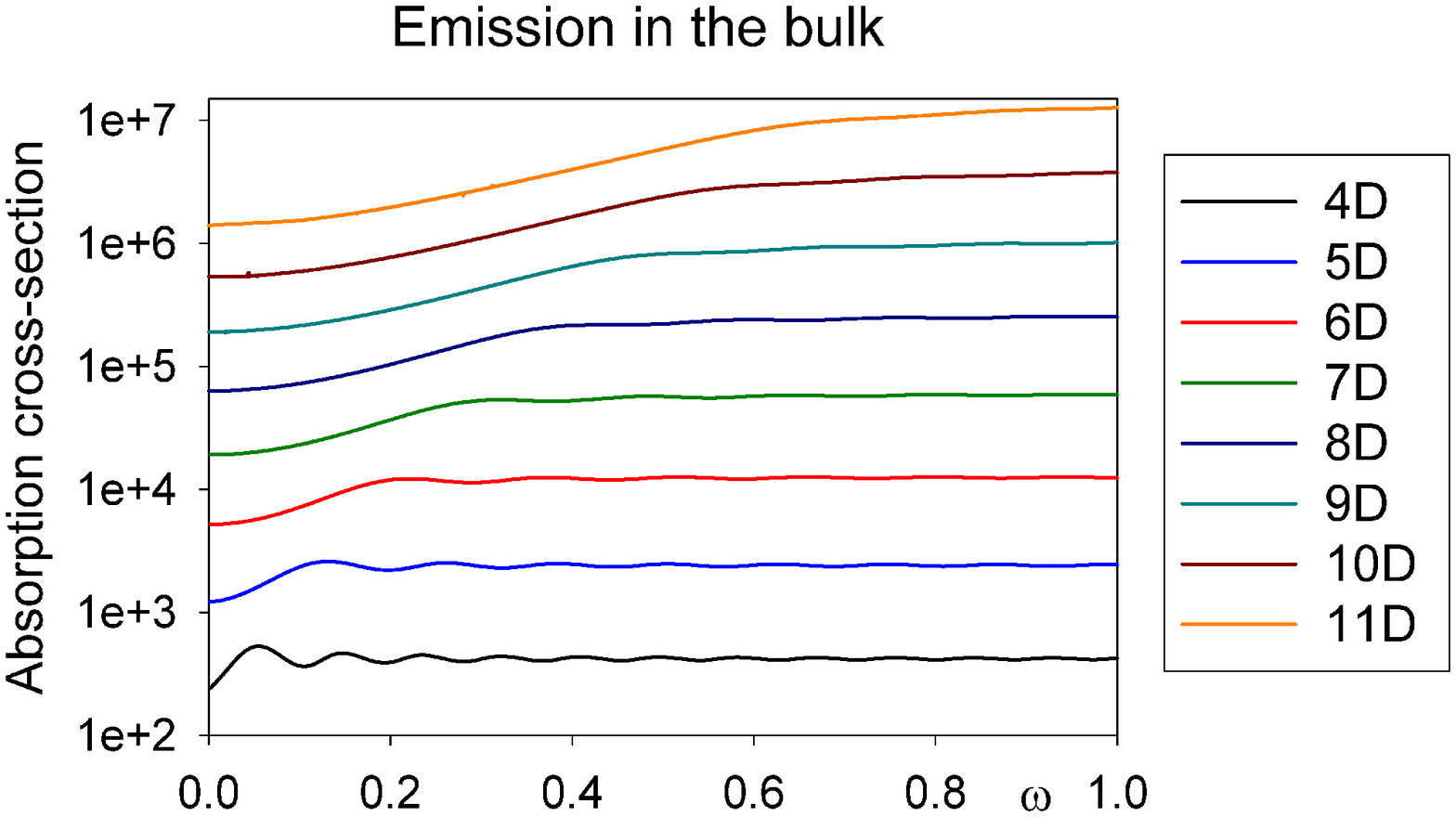}
\end{center}
\caption{Absorption cross-section (\ref{eq:bulkabsorption}) for a scalar field in the bulk as a function of frequency, for non-commutative black holes with maximum temperatures in varying numbers of space-time dimensions.
 The curves, from bottom to top, are for $n=0$ up to $n=7$.  We use units in which $M_{\star}$ and $\theta $ are set equal to unity.  For $n>0$ this corresponds to energies measured in TeV, but for $n=0$ energies are measured in units of $10^{16}$ TeV.  The different units are used for the $n=0$ case to facilitate comparison with the $n>0$ cases.}
\label{fig:absorptionbulk}
\end{figure}

We begin our discussion of bulk emission by considering the  absorption cross-section $\sigma ^{\mathrm {bulk}}(\omega )$ (\ref{eq:bulkabsorption}).
The bulk absorption cross-section shares many qualitative features with that for brane emission.  For low-frequency waves, the absorption cross-section tends to the area of the event horizon, which for higher-dimensional black holes is given by
\begin{equation}
A_{h} = \frac {2\pi r_{h}^{n+2}\pi ^{\frac {\left( n+1 \right) }{2}}}{\Gamma
\left( \frac {n+3}{2} \right) } .
\end{equation}
For five-dimensional black holes (figure~\ref{fig:5dabsorptionbulk}), the event horizon radius increases as the mass of the black hole increases (but is always smaller for NCBHs than for
Schwarzschild black holes with the same mass).
In figure~\ref{fig:absorptionbulk}, the area of the event horizon increases dramatically as the number of extra dimensions increases, and this
leads to the large increase in the absorption cross-section.

As the frequency increases, the absorption cross-section oscillates about its final, high-frequency limit, although the oscillations are of small amplitude for more than two extra dimensions.
In the high-frequency limit,
the absorption cross-section tends towards the projected area of an absorptive body of effective radius $r_{c}$  (\ref{eq:rc}).
The effective radius $r_{c}$ is the same for both bulk and brane modes \cite{Emparan:2000rs}, but care is needed in computing the relevant projected area (for a detailed discussion, see \cite{Harris:2003eg}).
The expected limiting behaviour of the absorption cross-section is \cite{Harris:2003eg}:
\begin{equation}
\sigma (\omega \rightarrow \infty ) =
\frac {2\pi ^{1+\frac {n}{2}}}{\left( n +2 \right) \Gamma \left( \frac {n+2}{2} \right) } \left( \frac {n+3}{2} \right) ^{\frac {n+2}{n+1}}
\left( \frac {n+3}{n+1} \right) ^{\frac {n+2}{2}} r_{h}^{n+2}.
\label{eq:highomegabulk}
\end{equation}
Comparing this formula with our numerical results in figures~\ref{fig:5dabsorptionbulk} and \ref{fig:absorptionbulk}, we find excellent
agreement by taking $r_{h}$ in (\ref{eq:highomegabulk}) to be the event horizon
radius of a Schwarzschild black hole having the same mass as the NCBH, except for larger numbers of extra dimensions, where the absorption cross-sections in figure~\ref{fig:absorptionbulk} have not yet converged to their asymptotic  limit.
Therefore, in the bulk, as on the brane, for high frequency waves the NCBHs are mimicking Schwarzschild black holes of the same mass.

\begin{figure}
\begin{center}
\includegraphics[width=10cm]{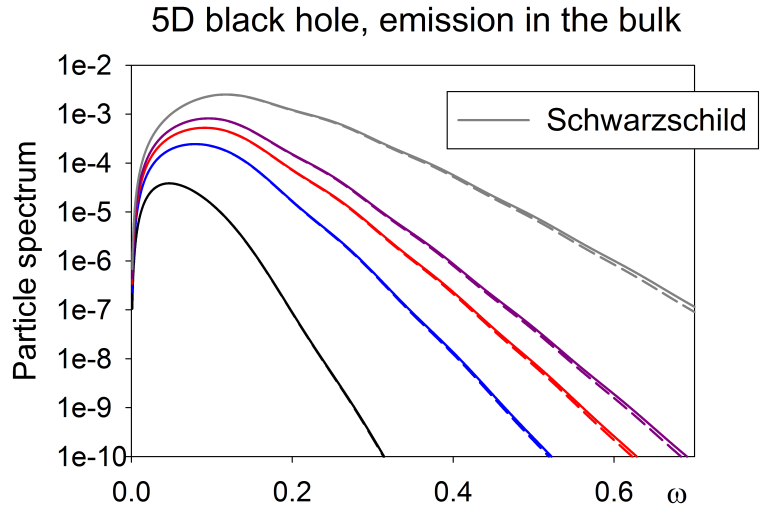}
\end{center}
\caption{Particle fluxes (\ref{eq:bulkparticle}, \ref{eq:bulkparticleNC}) for scalar field emission in the bulk, as a function of frequency, for
five-dimensional NCBHs with varying masses and a Schwarzschild black hole having the same mass as the NCBH with maximum temperature.
Solid lines indicate the standard particle flux (\ref{eq:bulkparticle}), and dotted lines the particle flux with additional damping due to non-commutative effects (\ref{eq:bulkparticleNC}).
The top curve (dark grey) is the flux for the Schwarzschild black hole, with the other curves being for NCBHs with increasing temperature from the bottom black curve to the top purple curve.  We use units in which $M_{\star}$ and $\theta $ are set equal to unity, corresponding to energies measured in TeV.}
\label{fig:5dbulkparticle}
\end{figure}
\begin{figure}
\begin{center}
\includegraphics[width=10cm]{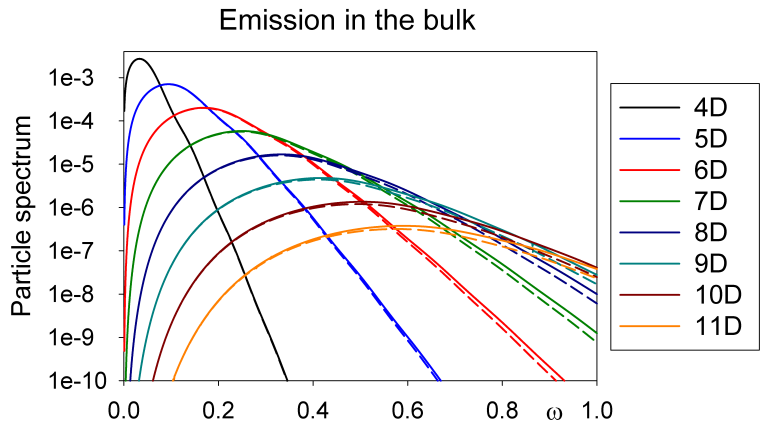}
\end{center}
\caption{Particle fluxes (\ref{eq:bulkparticle}, \ref{eq:bulkparticleNC}) for scalar field emission in the bulk, as a function of frequency, for NCBHs with maximum temperature and varying numbers of space-time dimensions.
Solid lines indicate the standard particle flux (\ref{eq:bulkparticle}), and dotted lines the particle flux with additional damping due to non-commutative effects (\ref{eq:bulkparticleNC}).
 The curves, from top to bottom, are for $n=0$ up to $n=7$.  We use units in which $M_{\star}$ and $\theta $ are set equal to unity.  For $n>0$ this corresponds to energies measured in TeV, but for $n=0$ energies are measured in units of $10^{16}$ TeV.  The different units are used for the $n=0$ case to facilitate comparison with the $n>0$ cases.}
\label{fig:bulkparticle}
\end{figure}

\begin{figure}
\begin{center}
\includegraphics[width=10cm]{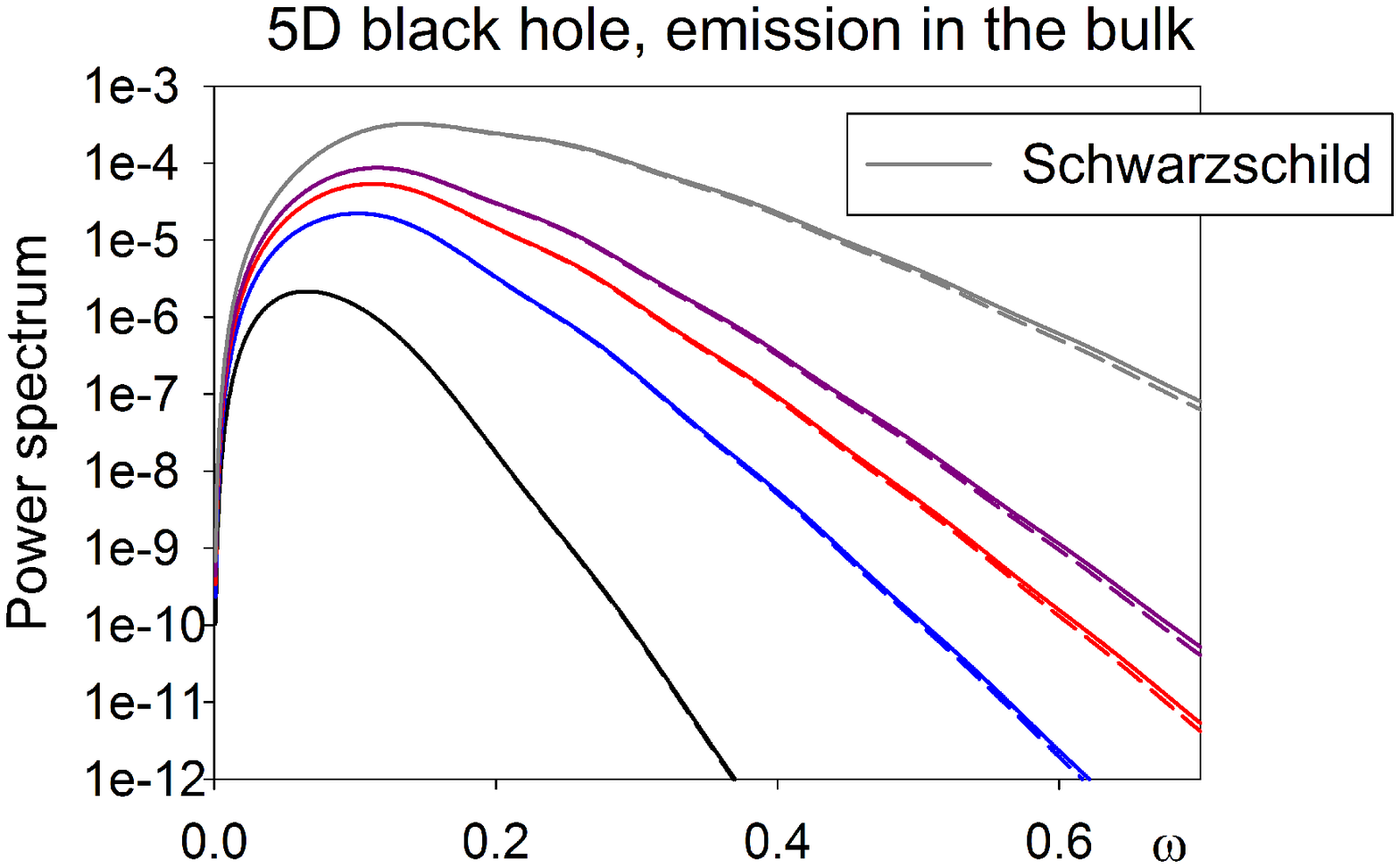}
\end{center}
\caption{Energy fluxes (\ref{eq:bulkpower}, \ref{eq:bulkpowerNC}) for scalar field emission in the bulk, as a function of frequency. The same black holes are considered as in figure~\ref{fig:5dbulkparticle}.
As in figure~\ref{fig:5dbulkparticle}, solid lines correspond to the standard energy flux (\ref{eq:bulkpower}) and dotted lines the energy flux (\ref{eq:bulkpowerNC})
with additional damping due to non-commutative effects.
The top curve (dark grey) is the flux for the Schwarzschild black hole, with the other curves being for NCBHs with increasing temperature from the bottom black curve to the top purple curve.  We use units in which $M_{\star}$ and $\theta $ are set equal to unity, corresponding to energies measured in TeV.}
\label{fig:5dbulkpower}
\end{figure}
\begin{figure}
\begin{center}
\includegraphics[width=11cm]{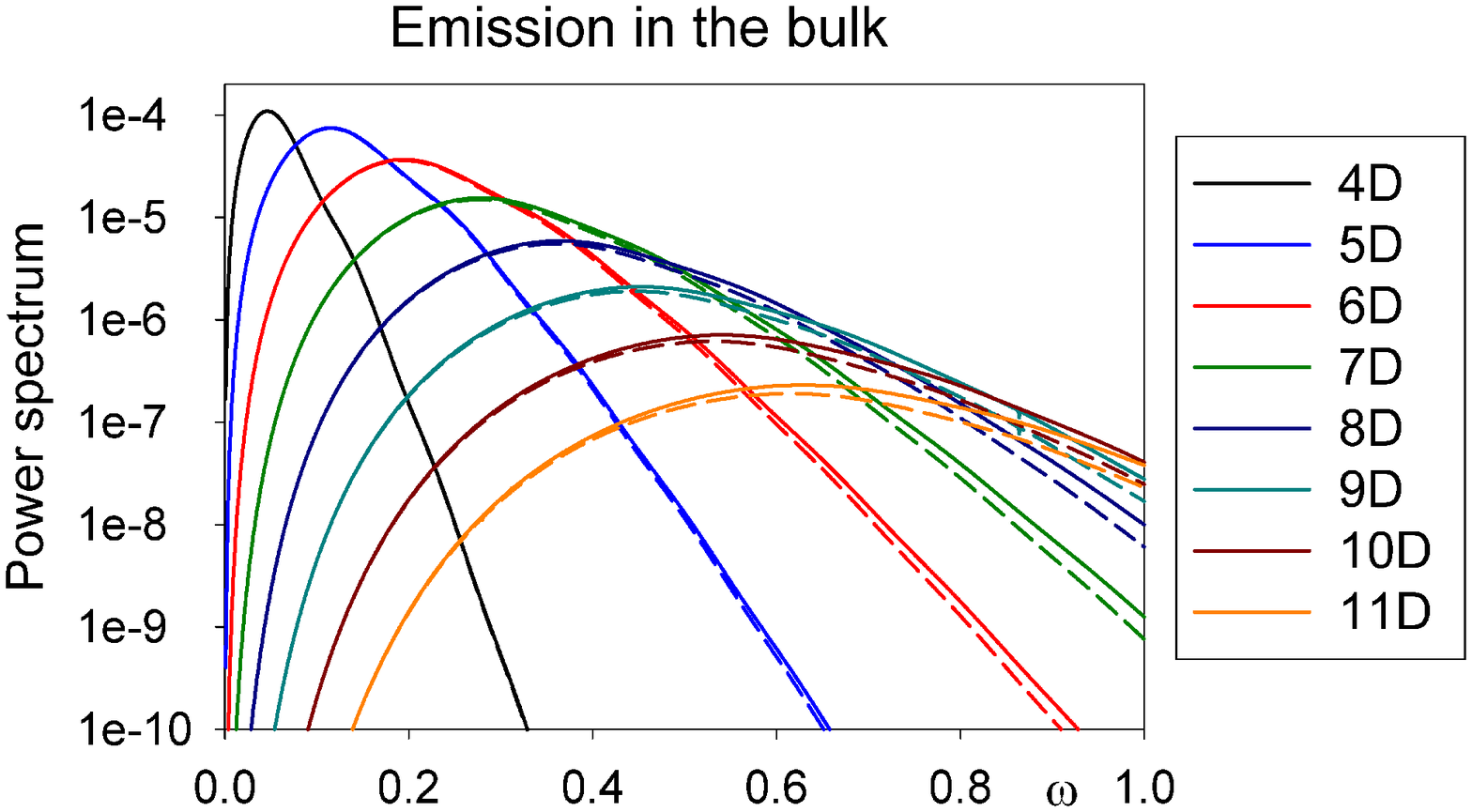}
\end{center}
\caption{Energy fluxes (\ref{eq:bulkpower}, \ref{eq:bulkpowerNC}) for scalar field emission in the bulk, as a function of frequency. The same black holes are considered as in figure \ref{fig:bulkparticle}.
As in figure \ref{fig:bulkparticle}, solid lines correspond to the standard energy flux (\ref{eq:bulkpower}) and dotted lines the energy flux (\ref{eq:bulkpowerNC})
with additional damping due to non-commutative effects.
 The curves, from top to bottom, are for $n=0$ up to $n=7$.  We use units in which $M_{\star}$ and $\theta $ are set equal to unity.  For $n>0$ this corresponds to energies measured in TeV, but for $n=0$ energies are measured in units of $10^{16}$ TeV.  The different units are used for the $n=0$ case to facilitate comparison with the $n>0$ cases.}
\label{fig:bulkpower}
\end{figure}
We now consider the particle
and energy fluxes (\ref{eq:bulkparticle}--\ref{eq:bulkpowerNC}), see figures~\ref{fig:5dbulkparticle}--\ref{fig:bulkpower}.
The results for particle and energy emission are similar.  For five-dimensional black holes (figures~\ref{fig:5dbulkparticle} and \ref{fig:5dbulkpower}), as with the emission on the brane, the emission in the bulk is much smaller for the NCBHs than it is for Schwarzschild black holes.  This dominant effect is due to the smaller temperature of the NCBHs. The additional damping term present in (\ref{eq:bulkparticleNC}--\ref{eq:bulkpowerNC}) once again has a negligible effect on the emission.

As we increase the number of extra dimensions (figures~\ref{fig:bulkparticle} and
\ref{fig:bulkpower}), the results are very different from those obtained for Schwarzschild black holes \cite{Harris:2003eg}.  In the latter case the bulk emission increases greatly as the number of space-time dimensions increases,  due mostly to the linear increase of the black hole temperature with $n$ for fixed horizon radius.  In our case, the maximum temperature of the black holes does increase with $n$, but not so quickly as for Schwarzschild black holes (see figure~\ref{fig:temp}), and the temperature of the NCBHs is so low that the peak of the emission of both particles and energy decreases as $n$ increases.  The emission spectrum broadens as $n$ increases, with emission at higher frequencies making a more significant contribution to the total.
We also observe that the additional damping term in the spectrum due to non-commutativity (\ref{eq:bulkparticleNC}--\ref{eq:bulkpowerNC}) becomes more important as $n$ increases, although, even for $n=11$ the difference between the spectra with the additional damping and the spectra without the additional damping is not great, because of the low temperature of the black holes.

\subsection{Comparison of bulk/brane emission}
\label{sec:bulk/brane}

We now consider the total emission from the NCBHs and the proportion of this emission which is in the bulk space-time.
We begin by comparing the total emission of particles and energy, both on the brane and in the bulk, from NCBHs with maximum temperature.
The totals for emission frequencies up to $\omega = 1$ are presented in Tables \ref{tab:brane} and \ref{tab:bulk}, where the fluxes have been rescaled so that the flux from a four-dimensional NCBH with maximum temperature is equal to unity.
\begin{table}[h]
\begin{center}
\resizebox{\textwidth}{!}{
\begin{tabular}{|c||c|c|c|c|c|c|c|c|}
\hline
 & $n=0$ & $n=1$ & $n=2$ & $n=3$ & $n=4$ & $n=5$ & $n=6$ & $n=7$
\\
\hline
\hline
Particles (undamped) &
1 & 3.3 & 6.0 & 8.5 & 10.7 & 12.4 & 13.8 & 15.0
\\
\hline
Particles (damped) &
1 & 3.3 & 6.0 & 8.4  & 10.6 & 12.3 & 13.7 & 14.8
\\
\hline
Power (undamped) &
1 & 5.5 & 13.1 & 21.8 & 30.2 & 37.6 & 44.0 & 49.3
\\
\hline
Power (damped) &
1 & 5.5 & 13.0 & 21.5 & 30.0 & 36.8 & 42.9 & 48.0
\\
\hline
\end{tabular}
}
\end{center}
\caption{Total fluxes of particles and energy on the brane for frequencies up to $\omega = 1$, for non-commutative black holes with maximum temperature, compared with the emission from a four-dimensional non-commutative black hole.}
\label{tab:brane}
\end{table}
\begin{table}[h]
\begin{center}
\resizebox{\textwidth}{!}{
\begin{tabular}{|c||c|c|c|c|c|c|c|c|}
\hline
 & $n=0$ & $n=1$ & $n=2$ & $n=3$ & $n=4$ & $n=5$ & $n=6$ & $n=7$
\\
\hline
\hline
Particles (undamped) &
1 & 0.56 & 0.24 & 0.091 & 0.032 & 0.011 & 0.0033 & 0.0010
\\
\hline
Particles (damped) &
1 & 0.56 & 0.24 & 0.088 & 0.030 & 0.0095 & 0.0029 & 0.00083
\\
\hline
Power (undamped) &
1 & 1.46 & 1.08 & 0.59 & 0.27 & 0.11 & 0.042 & 0.014
\\
\hline
Power (damped) &
1 & 1.45 & 1.05 & 0.56 & 0.25 & 0.098 & 0.035 & 0.012
\\
\hline
\end{tabular}
}
\end{center}
\caption{Total fluxes of particles and energy in the bulk for frequencies up to $\omega = 1$, for non-commutative black holes with maximum temperature, compared with the emission from a four-dimensional non-commutative black hole.}
\label{tab:bulk}
\end{table}

From table~\ref{tab:brane}, it is clear that the total emission of particles and energy on the brane steadily increases as $n$ increases, due to the increased temperature of the black holes with increasing $n$.  This is also evident from the plots of the brane emission as a function of frequency in figures~\ref{fig:braneparticle} and
\ref{fig:branepower}.
For emission in the bulk, the results are rather different, in accordance with our earlier discussion of figures~\ref{fig:bulkparticle} and \ref{fig:bulkpower}.
The fluxes of particles decrease as $n$ increases, however, the flux of energy is larger for $n=1$ and $n=2$ than it is for $n=0$, but for $n\ge 3$ it decreases steadily as $n$ increases.
In both table~\ref{tab:brane} and \ref{tab:bulk}, it can be seen that the additional damping due to non-commutativity effects makes only a small difference to the total emission.

\begin{figure}
\begin{center}
\includegraphics[width=12cm]{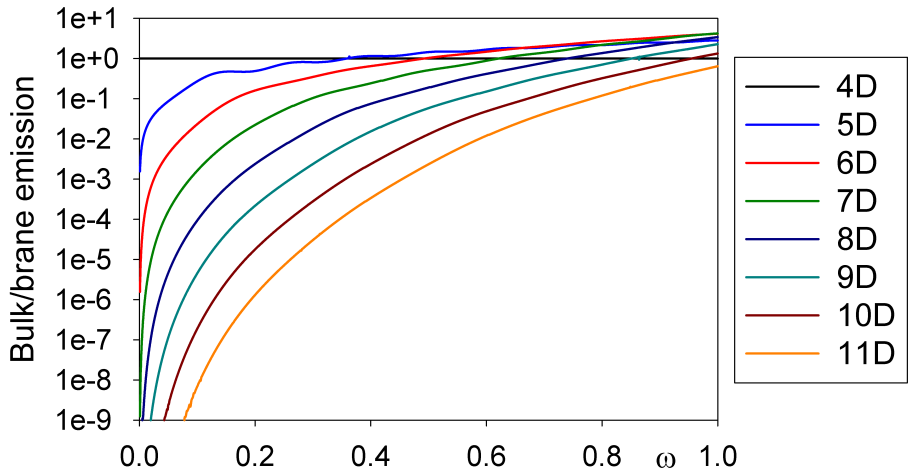}
\end{center}
\caption{Ratio of bulk/brane emission as a function of frequency, for NCBHs with maximum temperature. The curves, from top to bottom, are for $n=1$ to $n=7$ (the ratio for $n=0$ is unity for all $\omega $). The ratios are the same for particle and energy fluxes, and independent of whether there are additional damping terms in the spectra due to non-commutativity.}
\label{fig:bulkbraneratio}
\end{figure}

The ratio of bulk/brane emission is shown as a function of frequency $\omega $ in figure~\ref{fig:bulkbraneratio}.
It can be seen that the bulk emission is greatly suppressed compared to the brane emission for low frequencies and large $n$.  The bulk/brane ratio increases with frequency for all $n$, and for large $n$ becomes greater than unity for large frequencies.  The shape of the curves in figure~\ref{fig:bulkbraneratio} are qualitatively similar to those in \cite{Harris:2003eg} for Schwarzschild black holes, bearing in mind the different units we are using.
However, by comparing the values in table~\ref{tab:brane} and \ref{tab:bulk}, the ratio of total bulk emission to total brane emission can be seen to decrease rapidly as $n$ increases, from about $20\% $ bulk emission compared with brane emission for $n=1$ down to about $0.02\% $ bulk emission compared with brane emission for $n=7$.
This is in marked contrast to the results for Schwarzschild black holes \cite{Harris:2003eg}, where the bulk/brane ratio decreases down to about $22\% $ for $n=3$ but then increases as $n$ increases until it is about $93\% $ for $n=7$.
The reason why the bulk emission is so suppressed in our case is that the NCBHs have a very much smaller temperature compared with Schwarzschild black holes with the same mass.  Thus, while the bulk/brane ratio increases quickly with increasing frequency $\omega $ (see figure~\ref{fig:bulkbraneratio}), the low temperature means that there is negligible emission in high frequencies either on the brane or in the bulk, so that low-frequency emission dominates and this is mostly on the brane.

\section{Conclusions}
\label{sec:conc}

In the present paper we have addressed the problem of the Hawking emission from quantum gravity corrected black hole space-times.
One of the motivations for this study is recent LHC bounds on the fundamental scale of quantum gravity $M_{\star }$, namely
$M_{\star }\gtrsim 2-3 $ TeV \cite{Aad:2011xw,Chatrchyan:2011nd}, which implies that any black holes created at the LHC at energies up to 14 TeV cannot be adequately modelled as classical objects.

After reviewing the literature on the topic, we chose to start the analysis with the case of spherically symmetric NCBHs, whose modified thermodynamics encodes effects which are common to other quantum gravity modified black hole solutions.
We continued with a presentation of NCBHs and with an update about relevant quantities concerning their possible production in high energy hadron collisions at the LHC. We showed that NCBHs might be plentifully produced at the LHC or in the showers of cosmic rays hitting the upper layers of the Earth's atmosphere. Furthermore, it turned out that NCBHs are weakly affected by back-reaction effects and that their evaporation can be efficiently described by means of a semi-classical formalism.

We have performed a detailed study of the emission spectra for scalar fields on the brane and on the bulk. We showed that the simultaneous inclusion of a minimal length in both the geometry and the matter sector does not lead to significantly different results with respect to the case in which non-commutativity is present at the level of the background geometry only. The key feature which modifies the scenario with respect to the case of classical Schwarzschild black holes is the possibility of having a maximum temperature and a smaller horizon size at equivalent black hole mass. As a result we found reduced emission spectra with respect to the corresponding Schwarzschild black hole metrics. Even if for now we do not have a rigorous proof, we argue that this reduction is a model independent result, since it solely depends on the horizon extremization which is common to ASGBHs and to LQBHs too.

In addition, we found that NCBHs have striking differences in the bulk/brane emission ratio with respect to the classical metrics: the bulk emission drops to $0.02 \%$ of the brane emission as one increased the number $n$ of extra dimensions, while it increases with $n$ for the Schwarzschild case.
In other words, we find that the emission is dominated by low frequency modes, mostly on the brane.
This is the most phenomenologically interesting effect. The amount of energy lost in the bulk can be measured as missing energy by an observer on the brane. Such a missing energy determines the remaining available energy for emission on the brane in terms of easily detectable standard model particles. We stress that the observation of such peculiarities would not only confirm our predictions but might disclose further features about the nature of quantum gravity itself.

The work initiated in this paper is far from being concluded. The present analysis concerns just the black hole direct emission of scalar particles, while little is known about the subsequent evolution of matter and radiation.  For four-dimensional black holes, an increase of the spin of matter fields is responsible for a suppression of the emission since particles have to traverse a higher angular momentum barrier. The number of extra dimensions also affects the black hole emission, but in a specific manner for each value of the spin.
For example, for ten-dimensional, spherically symmetric, black holes, the emission of scalars, fermions and gauge bosons is comparable for each field degree of freedom \cite{Harris:2003eg}.
We cannot infer that the same pattern emerges when considering QGBHs.
The study of higher spin fields in QGBH backgrounds will be of primary importance, since it is directly connected to observations in particle detectors.
The emission of higher spin fields is also connected to the onset of the photosphere and the chromosphere, regions around the black hole where an electron-positron-photon plasma and a quark-gluon plasma might develop by means of particle production and bremsstrahlung mechanisms. To date quantum gravity effects have been neglected in studies of both the photosphere and chromosphere.

Another open direction of investigation is the evolution of black holes in phases preceding the spherically symmetric neutral configuration. For now we ignore how quantum gravity effects could affect black hole formation and the eventual loss of charge and/or angular momentum.
We expect a variety of new effects from the combination of higher spin fields, number of extra dimensions and non-spherical QGBH emission. In particular, it would be interesting to study the role of super-radiance in QGBH decay.
We plan to address these issues in forthcoming contributions.

\begin{acknowledgments}
This work is supported by the Helmholtz International Center for FAIR within the
framework of the LOEWE program (Landesoffensive zur Entwicklung Wissenschaftlich-\"{O}konom-ischer Exzellenz) launched by the State of Hesse.  This work is supported in part by the European Cooperation in Science and Technology (COST) action MP0905 ``Black Holes in a Violent Universe''. P.N. would like to thank the School of Mathematics and Statistics, University of Sheffield for the kind hospitality during a visit supported by COST action MP0905 at the initial stages of this work.
E.W. would like to thank the Frankfurt Institute for Advanced Studies, Goethe University Frankfurt for the kind hospitality during a visit during the final stage of this work.  E.W. also thanks Perimeter Institute, Waterloo, Canada and the School of Mathematical Sciences, Dublin City University for hospitality while this work was in progress; Sam Dolan for helpful discussions and Carl Kent for help with numerical computations.

\end{acknowledgments}

\providecommand{\href}[2]{#2}\begingroup\raggedright\endgroup


\begin{thebibliography}{10%
0}

\bibitem{Haw75}
S.~W. Hawking, {\it {Particle creation by black holes}},  {\em Commun. Math.
  Phys.} {\bf 43} (1975) 199--220.

\bibitem{GLAST}
\url{http://fermi.gsfc.nasa.gov/}.

\bibitem{Tho72}
K.~Thorne, {\em Nonspherical gravitational collapse, a short review.},
  pp.~231--258.
\newblock In J. Klauder et al., Magic Without Magic: J. A. Wheeler.
\newblock Freeman, San Francisco, 1972.

\bibitem{Meade:2007sz}
P.~Meade and L.~Randall, {\it {Black holes and quantum gravity at the {LHC}}},
  {\em JHEP} {\bf 05} (2008) 003,
  [\href{http://xxx.lanl.gov/abs/0708.3017}{{\tt arXiv:0708.3017}}].

\bibitem{DFG10}
G.~Dvali, S.~Folkerts, and C.~Germani, {\it {Physics of trans-{P}lanckian
  gravity}},  \href{http://xxx.lanl.gov/abs/1006.0984}{{\tt arXiv:1006.0984}}.

\bibitem{Spa11}
E.~Spallucci and S.~Ansoldi, {\it {Regular black holes in {UV} self-complete
  quantum gravity}},  {\em Phys. Lett.} {\bf B701} (2011) 471--474,
  [\href{http://xxx.lanl.gov/abs/1101.2760}{{\tt arXiv:1101.2760}}].

\bibitem{AADD98}
  I.~Antoniadis, N.~Arkani-Hamed, S.~Dimopoulos and G.~R.~Dvali,
  {\it{New dimensions at a millimeter to a {F}ermi and superstrings at a {T}e{V}}},
 {\em Phys.\ Lett.} {\bf B436} (1998) 257,
[\href{http://xxx.lanl.gov/abs/hep-ph/9804398}{{\tt hep-ph/9804398}}].

\bibitem{ADD99}
N.~Arkani-Hamed, S.~Dimopoulos, and G.~R. Dvali, {\it {Phenomenology,
  astrophysics and cosmology of theories with sub-millimeter dimensions and {T}e{V}
  scale quantum gravity}},  {\em Phys. Rev.} {\bf D59} (1999) 086004,
  [\href{http://xxx.lanl.gov/abs/hep-ph/9807344}{{\tt hep-ph/9807344}}].




\bibitem{RaS99}
L.~Randall and R.~Sundrum, {\it {A large mass hierarchy from a small extra
  dimension}},  {\em Phys. Rev. Lett.} {\bf 83} (1999) 3370--3373,
  [\href{http://xxx.lanl.gov/abs/hep-ph/9905221}{{\tt hep-ph/9905221}}].

\bibitem{DiL01}
S.~Dimopoulos and G.~L. Landsberg, {\it {Black holes at the {LHC}}},  {\em Phys.
  Rev. Lett.} {\bf 87} (2001) 161602,
  [\href{http://xxx.lanl.gov/abs/hep-ph/0106295}{{\tt hep-ph/0106295}}].

\bibitem{CMS11}
{\bf CMS} Collaboration, V.~Khachatryan {\em et.~al.}, {\it {Search for
  microscopic black hole signatures at the {L}arge {H}adron {C}ollider}},  {\em Phys.
  Lett.} {\bf B697} (2011) 434--453,
  [\href{http://xxx.lanl.gov/abs/1012.3375}{{\tt arXiv:1012.3375}}].

\bibitem{Cav03}
M.~Cavaglia, {\it {Black hole and brane production in {T}e{V} gravity: A review}},
  {\em Int. J. Mod. Phys.} {\bf A18} (2003) 1843--1882,
  [\href{http://xxx.lanl.gov/abs/hep-ph/0210296}{{\tt hep-ph/0210296}}].

\bibitem{Kan04}
P.~Kanti, {\it {Black holes in theories with large extra dimensions: A
  Review}},  {\em Int. J. Mod. Phys.} {\bf A19} (2004) 4899--4951,
  [\href{http://xxx.lanl.gov/abs/hep-ph/0402168}{{\tt hep-ph/0402168}}].

\bibitem{Hos05}
S.~Hossenfelder, {\em What black holes can teach us}, pp.~155--192.
\newblock Focus on black hole research.
\newblock Nova Science Publishers, 400 Oser Avenue, Suite 1600 Hauppauge, NY
  11788, 2005.

\bibitem{Web05}
B.~Webber, {\it {Black holes at accelerators}},
  \href{http://xxx.lanl.gov/abs/hep-ph/0511128}{{\tt hep-ph/0511128}}.

\bibitem{CaS06}
A.~Casanova and E.~Spallucci, {\it {{T}e{V} mini black hole decay at future
  colliders}},  {\em Class. Quant. Grav.} {\bf 23} (2006) R45--R62,
  [\href{http://xxx.lanl.gov/abs/hep-ph/0512063}{{\tt hep-ph/0512063}}].

\bibitem{Win07}
E.~Winstanley, {\it {Hawking radiation from rotating brane black holes}},
  \href{http://xxx.lanl.gov/abs/0708.2656}{{\tt arXiv:0708.2656}}.

\bibitem{BlN10}
M.~Bleicher and P.~Nicolini, {\it {Large extra dimensions and small black holes
  at the {LHC}}},  {\em J. Phys. Conf. Ser.} {\bf 237} (2010) 012008,
  [\href{http://xxx.lanl.gov/abs/1001.2211}{{\tt arXiv:1001.2211}}].

\bibitem{Kanti:2008eq}
P.~Kanti, {\it {Black holes at the {LHC}}},  {\em Lect. Notes Phys.} {\bf 769}
  (2009) 387--423, [\href{http://xxx.lanl.gov/abs/0802.2218}{{\tt arXiv:0802.2218}}].

\bibitem{Kanti:2009sz}
P.~Kanti, {\it {Brane-world black holes}},  {\em J. Phys. Conf. Ser.} {\bf 189}
  (2009) 012020, [\href{http://xxx.lanl.gov/abs/0903.2147}{{\tt arXiv:0903.2147}}].

\bibitem{Allahverdizadeh:2010hi}
M.~Allahverdizadeh, J.~Kunz, and F.~Navarro-Lerida, {\it {Charged rotating
  black holes in higher dimensions}},
  {\it {J. Phys. Conf. Ser. }} {\bf {314}} (2011) 012109,
  \href{http://xxx.lanl.gov/abs/1012.5052}{{\tt arXiv:1012.5052}}.

\bibitem{BCC06}
E.~Berti, V.~Cardoso, and M.~Casals, {\it {Eigenvalues and eigenfunctions of
  spin-weighted spheroidal harmonics in four and higher dimensions}},  {\em
  Phys. Rev.} {\bf D73} (2006) 024013,
  [\href{http://xxx.lanl.gov/abs/gr-qc/0511111}{{\tt gr-qc/0511111}}].

\bibitem{CKW06}
M.~Casals, P.~Kanti, and E.~Winstanley, {\it {Brane decay of a
  (4+n)-dimensional rotating black hole. II: Spin-1 particles}},  {\em JHEP}
  {\bf 02} (2006) 051, [\href{http://xxx.lanl.gov/abs/hep-th/0511163}{{\tt
  hep-th/0511163}}].

\bibitem{CDK07}
M.~Casals, S.~R. Dolan, P.~Kanti, and E.~Winstanley, {\it {Brane decay of a
  (4+n)-dimensional rotating black hole. III: Spin-1/2 particles}},  {\em JHEP}
  {\bf 03} (2007) 019, [\href{http://xxx.lanl.gov/abs/hep-th/0608193}{{\tt
  hep-th/0608193}}].

\bibitem{CDK08}
M.~Casals, S.~R. Dolan, P.~Kanti, and E.~Winstanley, {\it {Bulk emission of
  scalars by a rotating black hole}},  {\em JHEP} {\bf 06} (2008) 071,
  [\href{http://xxx.lanl.gov/abs/0801.4910}{{\tt arXiv:0801.4910}}].

\bibitem{Harris:2003eg}
C.~M. Harris and P.~Kanti, {\it {Hawking radiation from a (4+n)-dimensional
  black hole: Exact results for the {S}chwarzschild phase}},  {\em JHEP} {\bf 10}
  (2003) 014, [\href{http://xxx.lanl.gov/abs/hep-ph/0309054}{{\tt
  hep-ph/0309054}}].

\bibitem{AAS97}
S.~Ansoldi, A.~Aurilia, and E.~Spallucci, {\it {Hausdorff dimension of a
  quantum string}},  {\em Phys. Rev.} {\bf D56} (1997) 2352--2361,
  [\href{http://xxx.lanl.gov/abs/hep-th/9705010}{{\tt hep-th/9705010}}].

\pagebreak

\bibitem{AJL05}
J.~Ambjorn, J.~Jurkiewicz, and R.~Loll, {\it {Spectral dimension of the
  universe}},  {\em Phys. Rev. Lett.} {\bf 95} (2005) 171301,
  [\href{http://xxx.lanl.gov/abs/hep-th/0505113}{{\tt hep-th/0505113}}].

\bibitem{LaR05}
O.~Lauscher and M.~Reuter, {\it {Fractal spacetime structure in asymptotically
  safe gravity}},  {\em JHEP} {\bf 10} (2005) 050,
  [\href{http://xxx.lanl.gov/abs/hep-th/0508202}{{\tt hep-th/0508202}}].

\bibitem{Mod09}
L.~Modesto, {\it {Fractal structure of loop quantum gravity}},  {\em Class.
  Quant. Grav.} {\bf 26} (2009) 242002,
  [\href{http://xxx.lanl.gov/abs/0812.2214}{{\tt arXiv:0812.2214}}].

\bibitem{MoN10a}
L.~Modesto and P.~Nicolini, {\it {Spectral dimension of a quantum universe}},
  {\em Phys. Rev.} {\bf D81} (2010) 104040,
  [\href{http://xxx.lanl.gov/abs/0912.0220}{{\tt arXiv:0912.0220}}].

\bibitem{NiS11}
P.~Nicolini and E.~Spallucci, {\it {Un-spectral dimension and quantum spacetime
  phases}},  {\em Phys. Lett.} {\bf B695} (2011) 290--293,
  [\href{http://xxx.lanl.gov/abs/1005.1509}{{\tt arXiv:1005.1509}}].

\bibitem{NiN11}
P.~Nicolini and B.~Niedner, {\it {Hausdorff dimension of a particle path in a
  quantum manifold}},  {\em Phys. Rev.} {\bf D83} (2011) 024017,
  [\href{http://xxx.lanl.gov/abs/1009.3267}{{\tt arXiv:1009.3267}}].

\bibitem{MuS11}
J.~R. Mureika and D.~Stojkovic, {\it {Detecting vanishing dimensions via
  primordial gravitational wave astronomy}},  {\em Phys. Rev. Lett.} {\bf 106}
  (2011) 101101, [\href{http://xxx.lanl.gov/abs/1102.3434}{{\tt arXiv:1102.3434}}].

\bibitem{Cal11a}
G.~Calcagni, {\it {Gravity on a multifractal}},  {\em Phys. Lett.} {\bf B697}
  (2011) 251--253, [\href{http://xxx.lanl.gov/abs/1012.1244}{{\tt arXiv:1012.1244}}].

\bibitem{Cal11b}
G.~Calcagni, {\it {Geometry of fractional spaces}},
  \href{http://xxx.lanl.gov/abs/1106.5787}{{\tt arXiv:1106.5787}}.

\bibitem{Sca99}
F.~Scardigli, {\it {Generalized uncertainty principle in quantum gravity from
  micro-black hole gedanken experiment}},  {\em Phys. Lett.} {\bf B452} (1999)
  39--44, [\href{http://xxx.lanl.gov/abs/hep-th/9904025}{{\tt
  hep-th/9904025}}].

\bibitem{BoR99}
A.~Bonanno and M.~Reuter, {\it {Quantum gravity effects near the null black
  hole singularity}},  {\em Phys. Rev.} {\bf D60} (1999) 084011,
  [\href{http://xxx.lanl.gov/abs/gr-qc/9811026}{{\tt gr-qc/9811026}}].


\bibitem{Mod04}
L.~Modesto, {\it {Disappearance of black hole singularity in quantum gravity}},
   {\em Phys. Rev.} {\bf D70} (2004) 124009,
  [\href{http://xxx.lanl.gov/abs/gr-qc/0407097}{{\tt gr-qc/0407097}}].

\bibitem{Nic09}
P.~Nicolini, {\it {Noncommutative black holes, the final appeal to quantum
  gravity: a review}},  {\em Int. J. Mod. Phys.} {\bf A24} (2009) 1229--1308,
  [\href{http://xxx.lanl.gov/abs/0807.1939}{{\tt arXiv:0807.1939}}].

\bibitem{MMN11}
L.~Modesto, J.~W. Moffat, and P.~Nicolini, {\it {Black holes in an ultraviolet
  complete quantum gravity}},  {\em Phys. Lett.} {\bf B695} (2011) 397--400,
  [\href{http://xxx.lanl.gov/abs/1010.0680}{{\tt arXiv:1010.0680}}].

  \bibitem{Modesto11}
L.~Modesto, {\it {Super-renormalizable quantum gravity}},  \href{http://xxx.lanl.gov/abs/1107.2403}{{\tt arXiv:1107.2403}}.


\bibitem{Nic05}
P.~Nicolini, {\it {A model of radiating black hole in noncommutative
  geometry}},  {\em J. Phys.} {\bf A38} (2005) L631--L638,
  [\href{http://xxx.lanl.gov/abs/hep-th/0507266}{{\tt hep-th/0507266}}].

\bibitem{NSS06a}
P.~Nicolini, A.~Smailagic, and E.~Spallucci, {\it {The fate of radiating black
  holes in noncommutative geometry}},  {\em ESA Spec. Publ.} {\bf 637} (2006)
  11.1, [\href{http://xxx.lanl.gov/abs/hep-th/0507226}{{\tt hep-th/0507226}}].

\bibitem{NSS06b}
P.~Nicolini, A.~Smailagic, and E.~Spallucci, {\it {Noncommutative geometry
  inspired {S}chwarzschild black hole}},  {\em Phys. Lett.} {\bf B632} (2006)
  547--551, [\href{http://xxx.lanl.gov/abs/gr-qc/0510112}{{\tt
  gr-qc/0510112}}].

\bibitem{MKP07b}
Y.~S. Myung, Y.-W. Kim, and Y.-J. Park, {\it {Thermodynamics and evaporation of
  the noncommutative black hole}},  {\em JHEP} {\bf 02} (2007) 012,
  [\href{http://xxx.lanl.gov/abs/gr-qc/0611130}{{\tt gr-qc/0611130}}].

\bibitem{ANS07}
S.~Ansoldi, P.~Nicolini, A.~Smailagic, and E.~Spallucci, {\it {Noncommutative
  geometry inspired charged black holes}},  {\em Phys. Lett.} {\bf B645} (2007)
  261--266, [\href{http://xxx.lanl.gov/abs/gr-qc/0612035}{{\tt
  gr-qc/0612035}}].

\bibitem{BMS08}
  R.~Banerjee, B.~R.~Majhi, S.~Samanta,
  {\it{Noncommutative black hole thermodynamics}},
  {\em Phys.\ Rev.}\  {\bf D77} (2008) 124035,
  [\href{http://xxx.lanl.gov/abs/0801.3583}{{\tt arXiv:0801.3583}}].

\bibitem{BMM09}
  R.~Banerjee, B.~R.~Majhi, S.~K.~Modak,
 {\it{Noncommutative {S}chwarzschild black hole and area law}},
  {\em Class.\ Quant.\ Grav.}\  {\bf 26} (2009) 085010,
  [\href{http://xxx.lanl.gov/abs/0802.2176}{{\tt arXiv:0802.2176}}].

\bibitem{ABN09}
I.~Arraut, D.~Batic, and M.~Nowakowski, {\it {A non commutative model for a
  mini black hole}},  {\em Class. Quant. Grav.} {\bf 26} (2009) 245006,
  [\href{http://xxx.lanl.gov/abs/0902.3481}{{\tt arXiv:0902.3481}}].


\bibitem{NiS10}
P.~Nicolini and E.~Spallucci, {\it {Noncommutative geometry inspired dirty
  black holes}},  {\em Class. Quant. Grav.} {\bf 27} (2010) 015010,
  [\href{http://xxx.lanl.gov/abs/0902.4654}{{\tt arXiv:0902.4654}}].

\bibitem{SmS10}
A.~Smailagic and E.~Spallucci, {\it {`Kerrr' black hole: the lord of the
  string}},  {\em Phys. Lett.} {\bf B688} (2010) 82--87,
  [\href{http://xxx.lanl.gov/abs/1003.3918}{{\tt arXiv:1003.3918}}].

\bibitem{ABN10}
I.~Arraut, D.~Batic, and M.~Nowakowski, {\it {Maximal extension of the
  {S}chwarzschild spacetime inspired by noncommutative geometry}},  {\em J. Math.
  Phys.} {\bf 51} (2010) 022503, [\href{http://xxx.lanl.gov/abs/1001.2226}{{\tt
  arXiv:1001.2226}}].

\bibitem{BGM10}
R.~Banerjee, S.~Gangopadhyay, and S.~K. Modak, {\it {Voros product,
  noncommutative {S}chwarzschild black hole and corrected area law}},  {\em Phys.
  Lett.} {\bf B686} (2010) 181--187,
  [\href{http://xxx.lanl.gov/abs/0911.2123}{{\tt arXiv:0911.2123}}].

\bibitem{MuN11}
J.~R. Mureika and P.~Nicolini, {\it {Aspects of noncommutative
  (1+1)-dimensional black holes}}, {\em Phys. Rev.} {\bf D84} (2011) 044020,
  [\href{http://xxx.lanl.gov/abs/1104.4120}{{\tt arXiv:1104.4120}}].

\bibitem{NiT11}
P.~Nicolini and G.~Torrieri, {\it {The {H}awking-{P}age crossover in noncommutative
  anti-de {S}itter space}}, {\em JHEP} {\bf 08} (2011) 097,  [\href{http://xxx.lanl.gov/abs/1105.0188}{{\tt
  arXiv:1105.0188}}].

\bibitem{ACS01}
R.~J. Adler, P.~Chen, and D.~I. Santiago, {\it {The generalized uncertainty
  principle and black hole remnants}},  {\em Gen. Rel. Grav.} {\bf 33} (2001)
  2101--2108, [\href{http://xxx.lanl.gov/abs/gr-qc/0106080}{{\tt
  gr-qc/0106080}}].

\bibitem{AAL06}
G.~Amelino-Camelia, M.~Arzano, Y.~Ling, and G.~Mandanici, {\it {Black-hole
  thermodynamics with modified dispersion relations and generalized uncertainty
  principles}},  {\em Class. Quant. Grav.} {\bf 23} (2006) 2585--2606,
  [\href{http://xxx.lanl.gov/abs/gr-qc/0506110}{{\tt gr-qc/0506110}}].

\bibitem{MKP07a}
Y.~S. Myung, Y.-W. Kim, and Y.-J. Park, {\it {Black hole thermodynamics with
  generalized uncertainty principle}},  {\em Phys. Lett.} {\bf B645} (2007)
  393--397, [\href{http://xxx.lanl.gov/abs/gr-qc/0609031}{{\tt
  gr-qc/0609031}}].

\bibitem{CMP11}
B.~Carr, L.~Modesto, and I.~Premont-Schwarz, {\it {Generalized uncertainty
  principle and self-dual black holes}},
  \href{http://xxx.lanl.gov/abs/1107.0708}{{\tt arXiv:1107.0708}}.

\bibitem{Mod06}
L.~Modesto, {\it {Loop quantum black hole}},  {\em Class. Quant. Grav.} {\bf
  23} (2006) 5587--5602, [\href{http://xxx.lanl.gov/abs/gr-qc/0509078}{{\tt
  gr-qc/0509078}}].

\bibitem{Mod08}
L.~Modesto, {\it {Gravitational collapse in loop quantum gravity}},  {\em Int.
  J. Theor. Phys.} {\bf 47} (2008) 357--373,
  [\href{http://xxx.lanl.gov/abs/gr-qc/0610074}{{\tt gr-qc/0610074}}].

\bibitem{MoP09}
L.~Modesto and I.~Premont-Schwarz, {\it {Self-dual black holes in {LQG}: theory
  and phenomenology}},  {\em Phys. Rev.} {\bf D80} (2009) 064041,
  [\href{http://xxx.lanl.gov/abs/0905.3170}{{\tt arXiv:0905.3170}}].

\bibitem{Mod10}
L.~Modesto, {\it {Semiclassical loop quantum black hole}},  {\em Int. J. Theor.
  Phys.} {\bf 49} (2010) 1649--1683.

\bibitem{HMP10}
S.~Hossenfelder, L.~Modesto, and I.~Premont-Schwarz, {\it {A model for
  non-singular black hole collapse and evaporation}},  {\em Phys. Rev.} {\bf
  D81} (2010) 044036, [\href{http://xxx.lanl.gov/abs/0912.1823}{{\tt
  arXiv:0912.1823}}].

\bibitem{CaM10}
F.~Caravelli and L.~Modesto, {\it {Spinning loop black holes}},  {\em Class.
  Quant. Grav.} {\bf 27} (2010) 245022,
  [\href{http://xxx.lanl.gov/abs/1006.0232}{{\tt arXiv:1006.0232}}].

\bibitem{BoR00}
A.~Bonanno and M.~Reuter, {\it {Renormalization group improved black hole
  spacetimes}},  {\em Phys. Rev.} {\bf D62} (2000) 043008,
  [\href{http://xxx.lanl.gov/abs/hep-th/0002196}{{\tt hep-th/0002196}}].

\bibitem{BoR06}
A.~Bonanno and M.~Reuter, {\it {Spacetime structure of an evaporating black
  hole in quantum gravity}},  {\em Phys. Rev.} {\bf D73} (2006) 083005,
  [\href{http://xxx.lanl.gov/abs/hep-th/0602159}{{\tt hep-th/0602159}}].

\bibitem{BuK10}
T.~Burschil and B.~Koch, {\it {Renormalization group improved black hole
  space-time in large extra dimensions}},  {\em Zh. Eksp. Teor. Fiz.} {\bf 92}
  (2010) 219--225, [\href{http://xxx.lanl.gov/abs/0912.4517}{{\tt arXiv:0912.4517}}].

\bibitem{FLR10}
K.~Falls, D.~F. Litim, and A.~Raghuraman, {\it {Black holes and asymptotically
  safe gravity}},  \href{http://xxx.lanl.gov/abs/1002.0260}{{\tt arXiv:1002.0260}}.

\bibitem{CaE10}
Y.-F. Cai and D.~A. Easson, {\it {Black holes in an asymptotically safe gravity
  theory with higher derivatives}},  {\em JCAP} {\bf 1009} (2010) 002,
  [\href{http://xxx.lanl.gov/abs/1007.1317}{{\tt arXiv:1007.1317}}].

\bibitem{MbK05}
M.~R. Mbonye and D.~Kazanas, {\it {A non-singular black hole model as a
  possible end-product of gravitational collapse}},  {\em Phys. Rev.} {\bf D72}
  (2005) 024016, [\href{http://xxx.lanl.gov/abs/gr-qc/0506111}{{\tt
  gr-qc/0506111}}].

\bibitem{Myu09}
Y.~S. Myung, {\it {Thermodynamics of black holes in the deformed
{H}o\v{r}ava-{L}ifshitz gravity}},  {\em Phys. Lett.} {\bf B678} (2009) 127--130,
  [\href{http://xxx.lanl.gov/abs/0905.0957}{{\tt arXiv:0905.0957}}].

\bibitem{Par09}
M.-I. Park, {\it {The black hole and cosmological solutions in {IR} modified
  {H}orava gravity}},  {\em JHEP} {\bf 09} (2009) 123,
  [\href{http://xxx.lanl.gov/abs/0905.4480}{{\tt arXiv:0905.4480}}].

\bibitem{KiK10}
E.~Kiritsis and G.~Kofinas, {\it {On {H}orava-{L}ifshitz `black holes'}},  {\em
  JHEP} {\bf 01} (2010) 122, [\href{http://xxx.lanl.gov/abs/0910.5487}{{\tt
  arXiv:0910.5487}}].

\bibitem{GHS10}
P.~Gaete, J.~A. Helayel-Neto, and E.~Spallucci, {\it {Un-graviton corrections
  to the {S}chwarzschild black hole}},  {\em Phys. Lett.} {\bf B693} (2010)
  155--158, [\href{http://xxx.lanl.gov/abs/1005.0234}{{\tt arXiv:1005.0234}}].

\bibitem{Nic10}
P.~Nicolini, {\it {Entropic force, noncommutative gravity and ungravity}},
  {\em Phys. Rev.} {\bf D82} (2010) 044030,
  [\href{http://xxx.lanl.gov/abs/1005.2996}{{\tt arXiv:1005.2996}}].

\bibitem{MuS10}
J.~R. Mureika and E.~Spallucci, {\it {Vector unparticle enhanced black holes:
  exact solutions and thermodynamics}},  {\em Phys. Lett.} {\bf B693} (2010)
  129--133, [\href{http://xxx.lanl.gov/abs/1006.4556}{{\tt arXiv:1006.4556}}].


\bibitem{Ansoldi:2008jw}
  S.~Ansoldi,
{  \it{Spherical black holes with regular center: A review of existing models
  including a recent realization with {G}aussian sources}},
\href{http://xxx.lanl.gov/abs/0802.0330}{{\tt arXiv:0802.0330}}.

\bibitem{Riz06}
T.~G. Rizzo, {\it {Noncommutative inspired black holes in extra dimensions}},
  {\em JHEP} {\bf 09} (2006) 021,
  [\href{http://xxx.lanl.gov/abs/hep-ph/0606051}{{\tt hep-ph/0606051}}].


\bibitem{SSN09}
E.~Spallucci, A.~Smailagic, and P.~Nicolini, {\it {Non-commutative geometry
  inspired higher-dimensional charged black holes}},  {\em Phys. Lett.} {\bf
  B670} (2009) 449--454, [\href{http://xxx.lanl.gov/abs/0801.3519}{{\tt
  arXiv:0801.3519}}].

\bibitem{MoN10b}
L.~Modesto and P.~Nicolini, {\it {Charged rotating noncommutative black
  holes}},  {\em Phys. Rev.} {\bf D82} (2010) 104035,
  [\href{http://xxx.lanl.gov/abs/1005.5605}{{\tt arXiv:1005.5605}}].

\bibitem{CaN08}
R.~Casadio and P.~Nicolini, {\it {The decay-time of non-commutative micro-black
  holes}},  {\em JHEP} {\bf 11} (2008) 072,
  [\href{http://xxx.lanl.gov/abs/0809.2471}{{\tt arXiv:0809.2471}}].

\bibitem{Gin10}
D.~M. Gingrich, {\it {Noncommutative geometry inspired black holes in higher
  dimensions at the {LHC}}},  {\em JHEP} {\bf 05} (2010) 022,
  [\href{http://xxx.lanl.gov/abs/1003.1798}{{\tt arXiv:1003.1798}}].

\bibitem{Sny47a}
H.~S. Snyder, {\it {Quantized space-time}},  {\em Phys. Rev.} {\bf 71} (1947)
  38--41.

\bibitem{Lan97}
G.~Landi, {\em An introduction to noncommutative spaces and their geometries},
  vol.~51.
\newblock Lecture Notes in Physics. New Series m: Monographs, Springer-Verlag,
  Berlin, New York, 1997.

\bibitem{Mad00}
J.~Madore, {\it {An introduction to noncommutative geometry}},  {\em Lect.
  Notes Phys.} {\bf 543} (2000) 231--273.

\pagebreak

\bibitem{DoN01}
M.~R. Douglas and N.~A. Nekrasov, {\it {Noncommutative field theory}},  {\em
  Rev. Mod. Phys.} {\bf 73} (2001) 977--1029,
  [\href{http://xxx.lanl.gov/abs/hep-th/0106048}{{\tt hep-th/0106048}}].

\bibitem{Sza03}
R.~J. Szabo, {\it {Quantum field theory on noncommutative spaces}},  {\em Phys.
  Rept.} {\bf 378} (2003) 207--299,
  [\href{http://xxx.lanl.gov/abs/hep-th/0109162}{{\tt hep-th/0109162}}].

\bibitem{SmS04}
  A.~Smailagic and E.~Spallucci,
  {\it{Lorentz invariance, unitarity and {UV}-finiteness of {QFT} on noncommutative
  spacetime}},
{\em  J.\ Phys.\ A}  {\bf 37} (2004) 1
  [{\em Erratum-ibid.\  A} {\bf 37} (2004) 7169],
   [\href{http://xxx.lanl.gov/abs/hep-th/0406174}{{\tt hep-th/0406174}}].


\bibitem{SSN06}
E.~Spallucci, A.~Smailagic, and P.~Nicolini, {\it {Trace anomaly in quantum
  spacetime manifold}},  {\em Phys. Rev.} {\bf D73} (2006) 084004,
  [\href{http://xxx.lanl.gov/abs/hep-th/0604094}{{\tt hep-th/0604094}}].

\bibitem{BMM11a}
E.~Brown, R.~B. Mann, and L.~Modesto, {\it {Stability of self-dual black
  holes}},  {\em Phys. Lett.} {\bf B695} (2011) 376--383,
  [\href{http://xxx.lanl.gov/abs/1006.4164}{{\tt arXiv:1006.4164}}].

\bibitem{BMM11b}
E.~G. Brown, R.~B. Mann, and L.~Modesto, {\it {Mass inflation in the loop black
  hole}},  \href{http://xxx.lanl.gov/abs/1104.3126}{{\tt arXiv:1104.3126}}.

\bibitem{BaN10}
D.~Batic and P.~Nicolini, {\it {Fuzziness at the horizon}},  {\em Phys. Lett.}
  {\bf B692} (2010) 32--35, [\href{http://xxx.lanl.gov/abs/1001.1158}{{\tt
  arXiv:1001.1158}}].

\bibitem{BrM11}
E.~Brown and R.~B. Mann, {\it {Instability of the noncommutative geometry
  inspired black hole}},  {\em Phys. Lett.} {\bf B695} (2011) 440--445,
  [\href{http://xxx.lanl.gov/abs/1012.4787}{{\tt arXiv:1012.4787}}].

\bibitem{MaN11}
R.~B. Mann and P.~Nicolini, {\it {Cosmological production of noncommutative
  black holes}},  {\em Phys. Rev.} {\bf D84} (2011) 064014, [\href{http://xxx.lanl.gov/abs/1102.5096}{{\tt arXiv:1102.5096}}].

\bibitem{Man97}
R.~B. Mann, {\it {Black holes of negative mass}},  {\it Class. Quant. Grav.}
  {\bf 14} (1997) 2927--2930,
  [\href{http://xxx.lanl.gov/abs/gr-qc/9705007}{{\tt gr-qc/9705007}}].

\bibitem{HuM09}
V.~Husain and R.~B. Mann, {\it {Thermodynamics and phases in quantum gravity}},
   {\it Class. Quant. Grav.} {\bf 26} (2009) 075010,
  [\href{http://xxx.lanl.gov/abs/0812.0399}{{\tt arXiv:0812.0399}}].

\bibitem{Aad:2011xw}
  {\bf {ATLAS}} Collaboration, G.~Aad {\it et al.},
  {\it {Search for new phenomena with the monojet and missing transverse momentum signature using the {ATLAS} detector in sqrt(s) = 7 {T}e{V} proton-proton collisions}},
{\it {Phys. Lett.}} {\bf {B705}} (2011) 294--312,
  [\href{http://xxx.lanl.gov/abs/1106.5327}{{\tt arXiv:1106.5327}}].

\bibitem{Chatrchyan:2011nd}
  {\bf {CMS}} Collaboration, S.~Chatrchyan {\it et al.},
  {\it {Search for new physics with a mono-jet and missing transverse energy in pp
  collisions at sqrt(s) = 7 {T}e{V}}},
  \href{http://xxx.lanl.gov/abs/1106.4775}{{\tt arXiv:1106.4775}}.



\bibitem{MyP86}
  R.~C.~Myers, M.~J.~Perry,
\textit{Black holes in higher dimensional space-times,}
{\em  Annals Phys.}\  {\bf 172} (1986) 304.

\bibitem{CERN1}
\url{http://lhc-commissioning.web.cern.ch/lhc-commissioning/}.

\bibitem{Frost:2009cf}
J.~A. Frost {\em et.~al.}, {\it {Phenomenology of production and decay of
  spinning extra-dimensional black holes at hadron colliders}},  {\em JHEP}
  {\bf 10} (2009) 014, [\href{http://xxx.lanl.gov/abs/0904.0979}{{\tt
  arXiv:0904.0979}}].

\bibitem{Dai:2007ki}
D.-C. Dai {\em et.~al.}, {\it {Black{M}ax: A black-hole event generator with
  rotation, recoil, split branes and brane tension}},  {\em Phys. Rev.} {\bf
  D77} (2008) 076007, [\href{http://xxx.lanl.gov/abs/0711.3012}{{\tt
  arXiv:0711.3012}}].

\bibitem{Cardoso:2005mh}
V.~Cardoso, M.~Cavaglia, and L.~Gualtieri, {\it {Hawking emission of gravitons
  in higher dimensions: Non-rotating black holes}},  {\em JHEP} {\bf 02}
  (2006) 021, [\href{http://xxx.lanl.gov/abs/hep-th/0512116}{{\tt
  hep-th/0512116}}].

\bibitem{Cardoso:2005vb}
V.~Cardoso, M.~Cavaglia, and L.~Gualtieri, {\it {Black hole particle emission
  in higher-dimensional spacetimes}},  {\em Phys. Rev. Lett.} {\bf 96} (2006)
  071301, [\href{http://xxx.lanl.gov/abs/hep-th/0512002}{{\tt
  hep-th/0512002}}].

\bibitem{Cornell:2005ux}
A.~S. Cornell, W.~Naylor, and M.~Sasaki, {\it {Graviton emission from a
  higher-dimensional black hole}},  {\em JHEP} {\bf 02} (2006) 012,
  [\href{http://xxx.lanl.gov/abs/hep-th/0510009}{{\tt hep-th/0510009}}].

\bibitem{Creek:2006ia}
S.~Creek, O.~Efthimiou, P.~Kanti, and K.~Tamvakis, {\it {Graviton emission in
  the bulk from a higher-dimensional {S}chwarzschild black hole}},  {\em Phys.
  Lett.} {\bf B635} (2006) 39--49,
  [\href{http://xxx.lanl.gov/abs/hep-th/0601126}{{\tt hep-th/0601126}}].

\bibitem{Kanti:2002nr}
P.~Kanti and J.~March-Russell, {\it {Calculable corrections to brane black hole
  decay. I: The scalar case}},  {\em Phys. Rev.} {\bf D66} (2002) 024023,
  [\href{http://xxx.lanl.gov/abs/hep-ph/0203223}{{\tt hep-ph/0203223}}].

\bibitem{Kanti:2002ge}
P.~Kanti and J.~March-Russell, {\it {Calculable corrections to brane black hole
  decay. II: Greybody factors for spin 1/2 and 1}},  {\em Phys. Rev.} {\bf D67}
  (2003) 104019, [\href{http://xxx.lanl.gov/abs/hep-ph/0212199}{{\tt
  hep-ph/0212199}}].

\bibitem{Hod:2011zzb}
S.~Hod, {\it {Bulk emission by higher-dimensional black holes: Almost perfect
  blackbody radiation}},  {\em Class. Quant. Grav.} {\bf 28} (2011) 105016,
  [\href{http://xxx.lanl.gov/abs/1107.0797}{{\tt arXiv:1107.0797}}].

\bibitem{Casals:2009st}
M.~Casals, S.~R. Dolan, P.~Kanti, and E.~Winstanley, {\it {Angular profile of
  emission of non-zero spin fields from a higher-dimensional black hole}},
  {\em Phys. Lett.} {\bf B680} (2009) 365--370,
  [\href{http://xxx.lanl.gov/abs/0907.1511}{{\tt arXiv:0907.1511}}].

\bibitem{Creek:2007pw}
S.~Creek, O.~Efthimiou, P.~Kanti, and K.~Tamvakis, {\it {Scalar emission in the
  bulk in a rotating black hole background}},  {\em Phys. Lett.} {\bf B656}
  (2007) 102--111, [\href{http://xxx.lanl.gov/abs/0709.0241}{{\tt arXiv:0709.0241}}].

\bibitem{Creek:2007tw}
S.~Creek, O.~Efthimiou, P.~Kanti, and K.~Tamvakis, {\it {Greybody factors in a
  rotating black-hole background-II : fermions and gauge bosons}},  {\em Phys.
  Rev.} {\bf D76} (2007) 104013, [\href{http://xxx.lanl.gov/abs/0707.1768}{{\tt
  arXiv:0707.1768}}].

\bibitem{Creek:2007sy}
S.~Creek, O.~Efthimiou, P.~Kanti, and K.~Tamvakis, {\it {Greybody factors for
  brane scalar fields in a rotating black-hole background}},  {\em Phys. Rev.}
  {\bf D75} (2007) 084043, [\href{http://xxx.lanl.gov/abs/hep-th/0701288}{{\tt
  hep-th/0701288}}].

\bibitem{Duffy:2005ns}
G.~Duffy, C.~Harris, P.~Kanti, and E.~Winstanley, {\it {Brane decay of a
  (4+n)-dimensional rotating black hole: Spin-0 particles}},  {\em JHEP} {\bf
  09} (2005) 049, [\href{http://xxx.lanl.gov/abs/hep-th/0507274}{{\tt
  hep-th/0507274}}].

\bibitem{Harris:2005jx}
C.~M. Harris and P.~Kanti, {\it {Hawking radiation from a (4+n)-dimensional
  rotating black hole}},  {\em Phys. Lett.} {\bf B633} (2006) 106--110,
  [\href{http://xxx.lanl.gov/abs/hep-th/0503010}{{\tt hep-th/0503010}}].

\bibitem{Flachi:2008yb}
A.~Flachi, M.~Sasaki, and T.~Tanaka, {\it {Spin polarization effects in micro
  black hole evaporation}},  {\em JHEP} {\bf 05} (2009) 031,
  [\href{http://xxx.lanl.gov/abs/0809.1006}{{\tt arXiv:0809.1006}}].

\bibitem{Frolov:2002as}
V.~P. Frolov and D.~Stojkovic, {\it {Black hole radiation in the brane world
  and recoil effect}},  {\em Phys. Rev.} {\bf D66} (2002) 084002,
  [\href{http://xxx.lanl.gov/abs/hep-th/0206046}{{\tt hep-th/0206046}}].



\bibitem{Frolov:2002xf}
V.~P. Frolov and D.~Stojkovic, {\it {Quantum radiation from a 5-dimensional
  rotating black hole}},  {\em Phys. Rev.} {\bf D67} (2003) 084004,
  [\href{http://xxx.lanl.gov/abs/gr-qc/0211055}{{\tt gr-qc/0211055}}].

\bibitem{Ida:2002ez}
D.~Ida, K.-y. Oda, and S.~C. Park, {\it {Rotating black holes at future
  colliders: Greybody factors for brane fields}},  {\em Phys. Rev.} {\bf D67}
  (2003) 064025, [\href{http://xxx.lanl.gov/abs/hep-th/0212108}{{\tt
  hep-th/0212108}}].

\bibitem{Ida:2005ax}
D.~Ida, K.-y. Oda, and S.~C. Park, {\it {Rotating black holes at future
  colliders. II: Anisotropic scalar field emission}},  {\em Phys. Rev.} {\bf
  D71} (2005) 124039, [\href{http://xxx.lanl.gov/abs/hep-th/0503052}{{\tt
  hep-th/0503052}}].

\bibitem{Ida:2006tf}
D.~Ida, K.-y. Oda, and S.~C. Park, {\it {Rotating black holes at future
  colliders. III: Determination of black hole evolution}},  {\em Phys. Rev.}
  {\bf D73} (2006) 124022, [\href{http://xxx.lanl.gov/abs/hep-th/0602188}{{\tt
  hep-th/0602188}}].

\bibitem{Kanti:2009sn}
P.~Kanti, H.~Kodama, R.~A. Konoplya, N.~Pappas, and A.~Zhidenko, {\it {Graviton
  emission in the bulk by a simply rotating black hole}},  {\em Phys. Rev.}
  {\bf D80} (2009) 084016, [\href{http://xxx.lanl.gov/abs/0906.3845}{{\tt
  arXiv:0906.3845}}].

\bibitem{Doukas:2009cx}
J.~Doukas, H.~T. Cho, A.~S. Cornell, and W.~Naylor, {\it {Graviton emission
  from simply rotating {K}err-de {S}itter black holes: Transverse traceless tensor
  graviton modes}},  {\em Phys. Rev.} {\bf D80} (2009) 045021,
  [\href{http://xxx.lanl.gov/abs/0906.1515}{{\tt arXiv:0906.1515}}].

\pagebreak

\bibitem{Kanti:2010mk}
P.~Kanti and N.~Pappas, {\it {Emission of massive scalar fields by a
  higher-dimensional rotating black-hole}},  {\em Phys. Rev.} {\bf D82} (2010)
  024039, [\href{http://xxx.lanl.gov/abs/1003.5125}{{\tt arXiv:1003.5125}}].

\bibitem{Rogatko:2009jp}
M.~Rogatko and A.~Szyplowska, {\it {Massive fermion emission from higher
  dimensional black holes}},  {\em Phys. Rev.} {\bf D79} (2009) 104005,
  [\href{http://xxx.lanl.gov/abs/0904.4544}{{\tt arXiv:0904.4544}}].

\bibitem{Sampaio:2009tp}
M.~O.~P. Sampaio, {\it {Distributions of charged massive scalars and fermions
  from evaporating higher-dimensional black holes}},  {\em JHEP} {\bf 02}
  (2010) 042, [\href{http://xxx.lanl.gov/abs/0911.0688}{{\tt arXiv:0911.0688}}].

\bibitem{Sampaio:2009ra}
M.~O.~P. Sampaio, {\it {Charge and mass effects on the evaporation of
higher-dimensional rotating black holes}},  {\em JHEP} {\bf 10} (2009) 008,
  [\href{http://xxx.lanl.gov/abs/0907.5107}{{\tt arXiv:0907.5107}}].

\bibitem{Grain:2005my}
J.~Grain, A.~Barrau, and P.~Kanti, {\it {Exact results for evaporating black
  holes in curvature-squared {L}ovelock gravity: {G}auss-{B}onnet greybody
  factors}},  {\em Phys. Rev.} {\bf D72} (2005) 104016,
  [\href{http://xxx.lanl.gov/abs/hep-th/0509128}{{\tt hep-th/0509128}}].

\bibitem{Konoplya:2010vz}
R.~A. Konoplya and A.~Zhidenko, {\it {Long life of {G}auss-{B}onnet corrected black
  holes}},  {\em Phys. Rev.} {\bf D82} (2010) 084003,
  [\href{http://xxx.lanl.gov/abs/1004.3772}{{\tt arXiv:1004.3772}}].

\bibitem{Unruh:1976db}
W.~G. Unruh, {\it {Notes on black hole evaporation}},  {\em Phys. Rev.} {\bf
  D14} (1976) 870.

\bibitem{DeF08}
Y.~Decanini and A.~Folacci, {\it {Hadamard renormalization of the stress-energy
  tensor for a quantized scalar field in a general spacetime of arbitrary
  dimension}},  {\em Phys. Rev.} {\bf D78} (2008) 044025,
  [\href{http://xxx.lanl.gov/abs/gr-qc/0512118}{{\tt gr-qc/0512118}}].

\bibitem{Emparan:2000rs}
R.~Emparan, G.~T. Horowitz, and R.~C. Myers, {\it {Black holes radiate mainly
  on the brane}},  {\em Phys. Rev. Lett.} {\bf 85} (2000) 499--502,
  [\href{http://xxx.lanl.gov/abs/hep-th/0003118}{{\tt hep-th/0003118}}].

\bibitem{Muller}
C.~Muller, {\em Spherical harmonics}, vol.~17 of {\em Lecture Notes in
  Mathematics}.
\newblock Springer-Verlag, Berlin-Heidelberg, Germany, 1966.

\end{thebibliography}
    \end{document}